\documentclass[pre,showpacs,showkeys,preprintnumbers,amsmath,amssymb,superscriptaddress,twocolumn,nofootinbib]{revtex4}
\usepackage{graphicx}
\usepackage{amsfonts}
\usepackage{url}
\usepackage{epstopdf}
\usepackage{color}
\usepackage{bm}
\usepackage[colorlinks=true,linkcolor=blue,citecolor=red]{hyperref}%

\def\L{\mathcal L}
\def\N{\mathcal N}
\def\var{\mathrm{Var}}

\def\Ei{{\rm Ei}}

\def\x{\bm{x}}
\def\X{\bm{X}}

\def\ve{\varepsilon}

\def\Re{{\rm Re}}

\def\pa{\partial\Omega}

\def\E{{\mathbb E}}

\def\P{{\mathbb P}}
\def\R{{\mathbb R}}

\def\T{{\mathcal T}}
\def\L{{\mathcal L}}

\def\rrho{R}

\def\erf{\mathrm{erf}}
\def\erfc{\mathrm{erfc}}
\def\erfcx{\mathrm{erfcx}}

\begin{document}

\title{Depletion of Resources by a Population of Diffusing Species}

\author{Denis~S.~Grebenkov}
 \email{denis.grebenkov@polytechnique.edu}
\affiliation{
Laboratoire de Physique de la Mati\`{e}re Condens\'{e}e (UMR 7643), \\ 
CNRS -- Ecole Polytechnique, IP Paris, 91128 Palaiseau, France}

\date{\today}

\begin{abstract}
Depletion of natural and artificial resources is a fundamental problem
and a potential cause of economic crises, ecological catastrophes, and
death of living organisms.  Understanding the depletion process is
crucial for its further control and optimized replenishment of
resources.  In this paper, we investigate a stock depletion by a
population of species that undergo an ordinary diffusion and consume
resources upon each encounter with the stock.  We derive the exact
form of the probability density of the random depletion time, at which
the stock is exhausted.  The dependence of this distribution on the
number of species, the initial amount of resources, and the geometric
setting is analyzed.  Future perspectives and related open problems
are discussed.
\end{abstract}


%02.50.-r       (Probability theory, stochastic processes, and statistics)
%05.40.-a 	Fluctuation phenomena, random processes, noise, and Brownian motion
%02.70.Rr       (General statistical methods)
%05.10.Gg 	Stochastic analysis methods (Fokker-Planck, Langevin, etc.) 

%02.50.Ey 	Stochastic processes  (Probability theory, stochastic processes, and statistics)


\maketitle

\section{Introduction}

How long does it take to deplete a finite amount of resources?  This
fundamental question naturally appears in many aspects of our everyday
life and in various disciplines, including economics and ecology.  On
a global scale, it may concern renewable and non-renewable natural
resources such as water, oil, forests, minerals, food, as well as
extinction of wildlife populations or fish stocks
\cite{Mangel85,Wada10,Dirzo14}.  On a local scale, one may think of
depletion-controlled starvation of a forager due to the consumption of
environmental resources
\cite{Benichou14,Chupeau16,Benichou16,Chupeau17,Bhat17,Benichou18}
that poses various problems of optimal search and exploration
\cite{Viswanathan,Viswanathan99,Benichou11,Gueudre14}.  On even finer,
microscopic scale, the depletion of oxygen, glucose, ions, ATP
molecules and other chemical resources is critical for life and death
of individual cells \cite{Fitts94,Parekh97,Ha99,Clapham07}.  A
reliable characterization of the depletion time, i.e., the instance of
an economical crisis, an ecological catastrophe, or the death of a
forager or a cell due to resources extinction, is a challenging
problem, whose solution clearly depends on the considered depletion
process.

\begin{figure}
\begin{center}
\includegraphics[width=33mm]{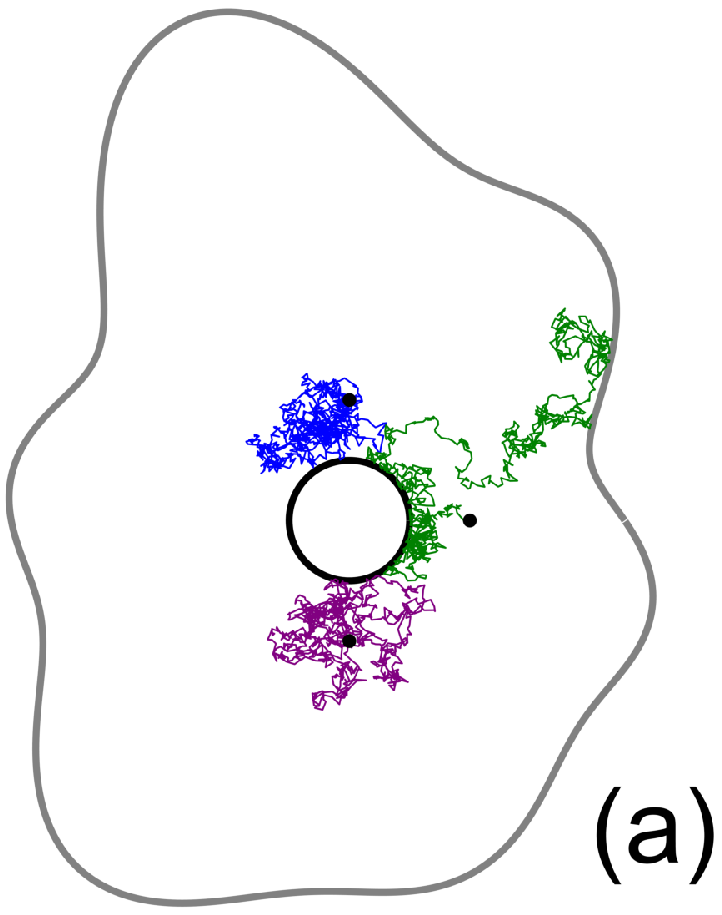}  % domain1.eps}
\includegraphics[width=52mm]{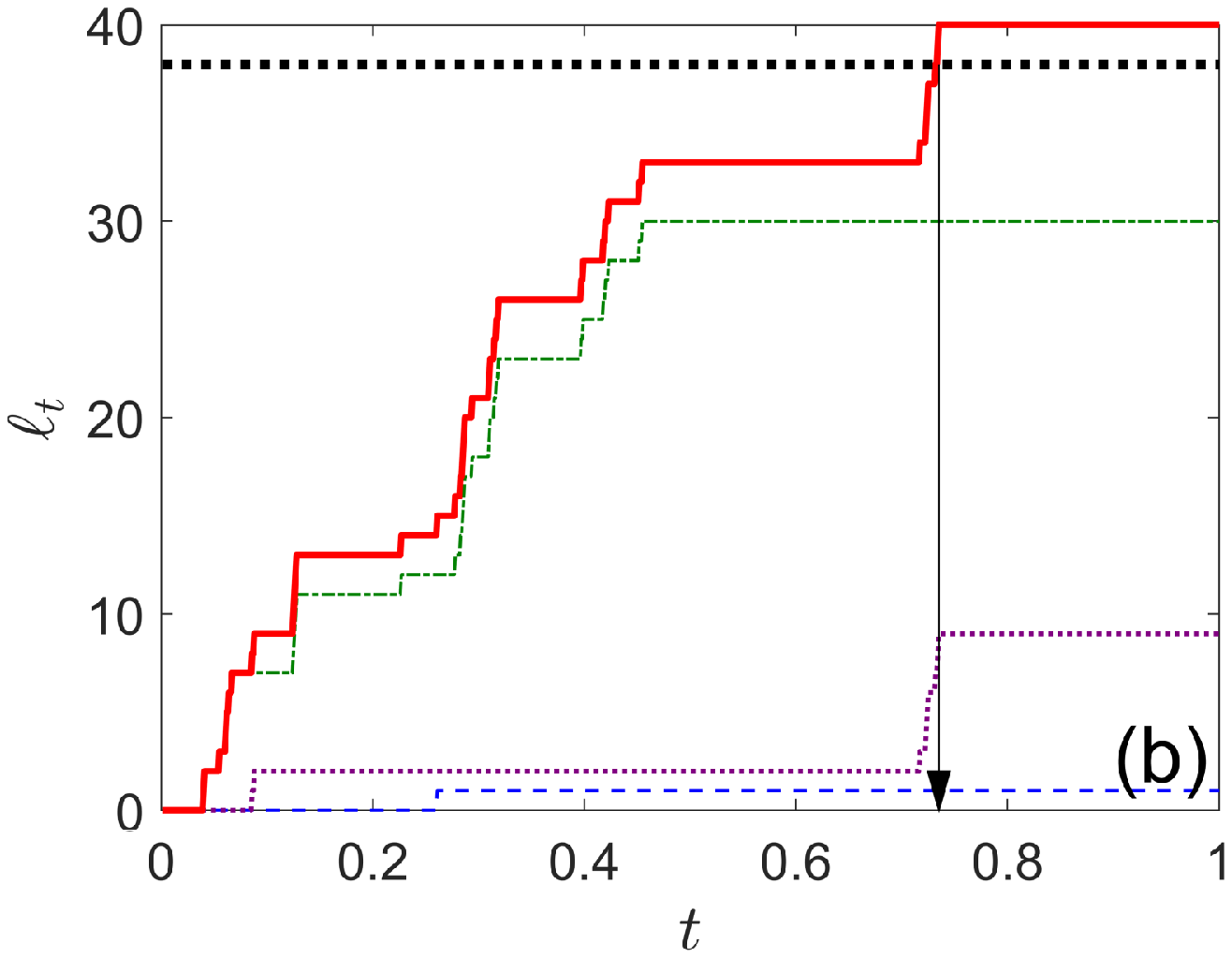}  % domain_ell1.eps}
\end{center}
\caption{
Schematic illustration of a stock depletion problem.  {\bf (a)} Random
trajectories of three species diffusing in a bounded domain with the
reflecting boundary (shown in gray); at each encounter with the stock
region (black circle), one unit of resources is consumed; here, the
species are released at different starting points (indicated by small
black disks) for a better visualization.  {\bf (b)} The number of
consumed resources (thick solid red curve), $\ell_t$, as a function of
time, and a prescribed threshold (thick dotted black horizontal line),
$\ell$, of initially available resources on the stock region; the
arrow indicates the first-crossing time $\T_{\ell,N}$ when the stock
is depleted.  Thin curves show the resources $\ell_t^i$ consumed by
individual species.  }
\label{fig:domain}
% load('domain_fig.mat');
% [an, Xt1,Xt2,Xt3, ell1,ell2,ell3] = A_localtime6_scheme_fig(an, Xt1,Xt2,Xt3, ell1,ell2,ell3);
\end{figure}

In this paper, we investigate a large class of stock depletion
processes inspired from biology and modeled as follows: there is a
population of $N$ independent species (or particles) searching for a
spatially localized stock of resources located on the impenetrable
surface of a bulk region (Fig. \ref{fig:domain}).  Any species that
has reached the location of the stock, receives a unit of resource and
continues its motion.  The species are allowed to return any number of
times to the stock, each time getting a unit of resource,
independently of its former delivery history and of other species.
This is a simple yet rich model of a diffusion-controlled release of
non-renewable resources upon request.  While the applicability of this
simplistic model for a quantitative description of natural depletion
phenomena is debatable, its theoretical analysis can reveal some
common, yet unexplored features of the general stock depletion
problem.

If the stock can be modeled as a node on a graph, which is accessed by
$N$ random walkers, the stock depletion problem is equivalent to
determining the first time when the total number of visits of that
site (or a group of sites) exceeds a prescribed threshold
\cite{Spitzer,Condamin05,Condamin07}.  In turn, for continuous-space
dynamics, two situations have to be distinguished: (i) The stock is a
bulk region, through which the species can freely diffuse; in this
case, each species is continuously receiving a fraction of resources
as long as it stays within the stock region; the total residence time
(also known as occupation or sojourn time) spent by $N$ species inside
the stock region can be considered as a proxy for the number of
released resources, an one is interested in the first time when this
total residence time exceeds a prescribed threshold.  The distribution
of the residence time for single and multiple particles has been
thoroughly investigated
\cite{Darling57,Ray63,Knight63,Agmon84,Berezhkovskii98,Dhar99,Yuste01,Godreche01,Majumdar02,Benichou03,Grebenkov07a,Burov07,Burov11}.
(ii) Alternatively, the stock can be located on the impenetrable
surface of a bulk region, in which case the species gets a unit of
resources at each encounter with that boundary region
(Fig. \ref{fig:domain}); the total number of encounters with the stock
region, which is a natural proxy for the number of released resources,
is characterized by the total boundary local time $\ell_t$ spent by
all species on the stock region
\cite{Levy,Ito,Grebenkov07a,Grebenkov19b,Grebenkov21a}.  In this
paper, we focus on this yet unexplored setting and aim at answering
the following question: If the amount of resources is limited, when
does the stock become empty?  The time of the stock depletion can be
formally introduced as the first-crossing time of a given threshold
$\ell$ (the initial amount of resources on the stock) by $\ell_t$:
\begin{equation}  \label{eq:T_def}
\T_{\ell,N} = \inf\{ t>0 ~:~ \ell_t > \ell\}.  
\end{equation}
We investigate the probability density of this random variable and its
dependence on the number $N$ of diffusing species, the initial amount
of resources $\ell$, and the geometric setting in which search occurs.
We also show how this problem generalizes the extreme first-passage
time statistics that got recently considerable attention
\cite{Weiss83,Basnayake19,Lawley20,Lawley20b,Lawley20c,Bray13,Majumdar20,Grebenkov20d}.

\section{Model and general solution}

We assume that $N$ independent point-like particles are released at
time $t = 0$ from a fixed starting point $\x_0 \in \Omega$ inside an
Euclidean domain $\Omega \subset \R^d$ with a smooth boundary $\pa$
(Fig. \ref{fig:domain}).  Each of these particles undertakes an
ordinary diffusion inside $\Omega$ with diffusion coefficient $D$ and
normal reflections on the impenetrable boundary $\pa$.  Let $\Gamma
\subset \pa$ denote a stock region (that we will also call a target),
on which resources are distributed.  For each particle $i$, we
introduce its boundary local time $\ell_t^i$ on the stock region
$\Gamma$ as $\ell_t^i = \lim\limits_{a\to 0} a \, \N_t^{a,i}$, where
$\N_t^{a,i}$ is the number of downcrossings of a thin boundary layer
of width $a$ near the stock region, $\Gamma_a = \{\x\in\Omega ~:~
|\x-\Gamma| < a\}$, up to time $t$
\cite{Levy,Ito,Grebenkov07a,Grebenkov19b,Grebenkov21a}.
In other words, $\N_t^{a,i}$ represents the number of encounters of
the $i$-th particle with the stock region $\Gamma$ (see
\cite{Grebenkov20} for further discussion).  While $\N_t^{a,i}$
diverges in the limit $a\to 0$ due to the self-similar nature of
Brownian motion, rescaling by $a$ yields a well-defined limit
$\ell_t^i$.  For a small width $a$, $\N_t^{a,i} \approx \ell_t^i/a$
can thus be interpreted as the number of resources consumed by the
$i$-th particle up to time $t$.  In the following, we deal directly
with the boundary local times $\ell_t^i$, which can be easily
translated into $\N_t^{a,i}$ for any small $a$.
 
For a single particle, the probability distribution of the random
process $\ell_t^i$ was studied in
\cite{Grebenkov07a,Grebenkov19b,Grebenkov21a}.  In particular, the
moment-generating function of $\ell_t^i$ was shown to be
\begin{equation}  \label{eq:ell_ti_exp}
\E_{\x_0} \{ e^{-q\ell_t^i} \} = S_q(t|\x_0) ,
\end{equation}
where $S_q(t|\x_0)$ is the survival probability, which satisfies the
(backward) diffusion equation
\begin{equation}  \label{eq:Sq_diff}
\partial_t S_q(t|\x_0) = D \Delta S_q(t|\x_0) \qquad (\x_0 \in \Omega), 
\end{equation}
with the initial condition $S_q(0|\x_0) = 1$ and the mixed
Robin-Neumann boundary condition:
\begin{subequations}  \label{eq:Sq_BC}
\begin{align}
\left. (\partial_n + q )S_q(t|\x_0)\right|_{\Gamma} & = 0,   \label{eq:Sq_BC_Robin}\\
\left. \partial_n S_q(t|\x_0) \right|_{\pa\backslash \Gamma} & = 0 
\end{align}
\end{subequations}
(for unbounded domains, the regularity condition $S_q(t|\x_0)\to 1$ as
$|\x_0|\to \infty$ is also imposed).  Here $\Delta$ is the Laplace
operator, and $\partial_n$ is the normal derivative at the boundary
oriented outward the domain $\Omega$.  The survival probability of a
diffusing particle in the presence of a partially reactive target has
been thoroughly investigated
\cite{Collins49,Sano79,Sano81,Shoup82,Zwanzig90,Sapoval94,Filoche99,Sapoval02,Grebenkov03,Berezhkovskii04,Grebenkov05,Grebenkov06a,Traytak07,Bressloff08,Lawley15,Galanti16,Lindsay17,Bernoff18b,Grebenkov17,Grebenkov19d,Guerin21}.
In particular, the parameter $q \geq 0$ characterizes the reactivity
of the target, ranging from an inert target for $q = 0$ to a perfect
sink or trap for $q = \infty$.  While we speak here about a reactive
target in the context of the survival probability, there is no
reaction in the stock depletion problem, in which the stock region is
inert.  In other words, we only explore the fundamental relation
(\ref{eq:ell_ti_exp}) between the survival probability and the
moment-generating function $\E_{\x_0} \{ e^{-q\ell_t^i} \}$ in order
to determine the probability density of the boundary local time
$\ell_t^i$ for a single particle, as well as the probability density
of the associated first-crossing time
\cite{Grebenkov20,Grebenkov20b}.

The amount of resources consumed up to time $t$ is modeled by the
total boundary local time,
\begin{equation}
\ell_t = \ell_t^1 + \ldots + \ell_t^N ,
\end{equation}
spent by all species on the stock region.  As the individual boundary
local times $\ell_t^i$ are independent, the moment-generating function
of $\ell_t$ reads
\begin{equation}  \label{eq:SqN}
\E_{\x_0}\{ e^{-q\ell_t}\} = \bigl(\E_{\x_0}\{ e^{-q\ell_t^1}\} \bigr)^N = \bigl[S_q(t|\x_0)\bigr]^N ,
\end{equation}
from which the probability density $\rho_N(\ell,t|\x_0)$ of $\ell_t$
is formally obtained via the inverse Laplace transform with respect to
$q$:
\begin{equation}  \label{eq:rhoN_ILq}
\rho_N(\ell,t|\x_0) = \L_{q,\ell}^{-1} \bigl\{ [S_q(t|\x_0)]^N \bigr\} .
\end{equation}
Since the total boundary local time is a non-decreasing process, the
cumulative distribution function of the first-crossing time
$\T_{\ell,N}$, defined by Eq. (\ref{eq:T_def}), is
\begin{equation}  \label{eq:QN_def}
Q_N(\ell,t|\x_0) = \P_{\x_0}\{ \T_{\ell,N} < t\} = \P_{\x_0}\{ \ell_t > \ell \} ,
\end{equation}
from which Eq. (\ref{eq:rhoN_ILq}) implies
\begin{equation}  \label{eq:QN_ILT}
Q_N(\ell,t|\x_0) = 1 - \L_{q,\ell}^{-1} \biggl\{\frac{[S_q(t|\x_0)]^N}{q}\biggr\} .
\end{equation}
In turn, the probability density of the first-crossing time is
obtained by time derivative:
\begin{equation}   \label{eq:UN_def}  
U_N(\ell,t|\x_0) = \partial_t Q_N(\ell,t|\x_0) = \L_{q,\ell}^{-1} \biggl\{ - \partial_t \frac{[S_q(t|\x_0)]^N}{q} \biggr\} .
\end{equation}
Equations (\ref{eq:QN_ILT}, \ref{eq:UN_def}) that fully characterize
the depletion time $\T_{\ell,N}$ in terms of the survival probability
$S_q(t|\x_0)$ of a single particle, present the first main result.

In the limit $\ell\to 0$, Eq. (\ref{eq:UN_def}) becomes
\begin{equation}  \label{eq:UN_fastest}
U_N(0,t|\x_0) = -\partial_t [S_\infty(t|\x_0)]^{N} ,
\end{equation}
i.e., we retrieved the probability density of the {\it fastest}
first-passage time among $N$ particles to a perfectly absorbing
target: $\T_{0,\ell} = \min\{\tau^1_\infty,\ldots,\tau^N_\infty\}$,
where $\tau^i_\infty = \inf\{t > 0~:~ \X_t^i \in \Gamma\}$ is the
first-passage time of the $i$-th particle to $\Gamma$
\cite{Weiss83,Basnayake19,Lawley20,Lawley20b,Lawley20c}.  
Our analysis thus extends considerably the topic of extreme
first-passage time statistics beyond the first arrival.
%
%While we focused on the first-crossing time of a fixed threshold
%$\ell$, the probability density $U_N(\ell,t|\x_0)$ can be related to
%other first-passage times.  
More generally, replacing a fixed threshold $\ell$ by a random
threshold $\hat{\ell}$ allows one to implement partially reactive
targets and various surface reaction mechanisms \cite{Grebenkov20}.
For instance, if $\hat{\ell}$ is an exponentially distributed variable
with mean $1/q$, i.e., $\P\{ \hat{\ell} > \ell\} = e^{-q\ell}$, then
the probability density of the first-crossing time $\T_{\hat{\ell},N}$
of the random threshold $\hat{\ell}$ is obtained by averaging
$U_N(\ell,t|\x_0)$ with the density $q e^{-q\ell}$ of $\hat{\ell}$
that yields according to Eq. (\ref{eq:UN_def}):
\begin{equation}
\int\limits_0^\infty d\ell \, q e^{-q\ell} \, U_N(\ell,t|\x_0) = - \partial_t \bigl[S_q(t|\x_0)\bigr]^N .
\end{equation}
One can notice that the right-hand side is precisely the probability
density of the minimum of $N$ independent first-passage times,
$\tau^1_q, \ldots, \tau^N_q$, to a partially reactive target with
reactivity parameter $q$.  In other words, we conclude that
\begin{equation}  \label{eq:Tmin_hat}
\T_{\hat{\ell},N} = \min\{ \tau^1_q,\ldots,\tau^N_q\} .
\end{equation}
In turn, the individual first-passage times can also be defined by
using the associated boundary local times as $\tau^i_q = \inf\{ t> 0
~:~ \ell_t^i > \hat{\ell}^i\}$, where $\hat{\ell}^1, \ldots,
\hat{\ell}^N$ are independent exponential random variables with the
mean $1/q$ \cite{Grebenkov20}.  Interestingly, while every $\tau^i_q$
is defined as the time of the first crossing of a random threshold
$\hat{\ell}^i$ by $\ell_t^i$ independently from each other, their
minimum can be defined via Eq. (\ref{eq:Tmin_hat}) as the first
crossing of the total boundary local time of a random threshold
$\hat{\ell}$ with the same $q$.

%(see section
%\ref{sec:extensions} for further
%details; in addition, we discuss in \ref{sec:general} some general
%properties of $Q_N(\ell,t|\x_0)$, in particular, its long-time limit
%$Q_N(\ell,\infty|\x_0)$ that determines the probability of the stock
%depletion).

While the above extension to multiple particles may look simple,
getting the actual properties of the probability density
$U_N(\ell,t|\x_0)$ is challenging.  In fact, the survival probability
$S_q(t|\x_0)$ depends on $q$ {\it implicitly}, through the Robin
boundary condition (\ref{eq:Sq_BC_Robin}), except for a few cases (see
two examples in Appendices \ref{sec:half-line} and \ref{sec:ball}).
%Even for an interval with partially reactive
%endpoints, the spectral expansion of $S_q(t|\x_0)$ involves the
%eigenvalues of the Laplace operator, which are obtained via a
%numerical solution of a transcendental trigonometric equation that
%keeps the dependence on $q$ implicit.  In this light, even a numerical
%computation of the density $U_N(\ell,t|\x_0)$ via the inverse Laplace
%transform is not elementary (see \secnumerical).  At the same time,
In the following, we first describe some general properties and then
employ Eq. (\ref{eq:UN_def}) to investigate the short-time and
long-time asymptotic behaviors of the probability density
$U_N(\ell,t|\x_0)$ to provide a comprehensive view onto the depletion
stock problem.

\subsection{General properties}
\label{sec:general}

Let us briefly discuss several generic properties of the cumulative
distribution function $Q_N(\ell,t|\x_0)$.  Since the total boundary
local time is a non-decreasing process, the time of crossing a higher
threshold is longer than the time of crossing a lower threshold.  In
probabilistic terms, this statement reads
\begin{equation}  \label{eq:QN_ineq}
Q_N(\ell_1,t|\x_0) \geq Q_N(\ell_2,t|\x_0)  \qquad (\ell_1 < \ell_2).
\end{equation}
In particular, setting $\ell_1 =0$ in this inequality yields an upper
bound for the cumulative distribution function:
\begin{equation}  \label{eq:QN_bound}
1 - [S_\infty(t|\x_0)]^N = Q_N(0,t|\x_0) \geq Q_N(\ell,t|\x_0) ,
\end{equation}
where we used the asymptotic behavior of Eq. (\ref{eq:QN_ILT}) as
$\ell\to 0$.  In the same vein, as the total boundary local time
$\ell_t$ is the sum of non-negative boundary local times $\ell_t^i$,
the cumulative distribution function monotonously increases with $N$:
\begin{equation}  \label{eq:QN_ineq2}
Q_{N_1}(\ell,t|\x_0) \leq Q_{N_2}(\ell,t|\x_0)  \qquad (N_1 < N_2).
\end{equation}

Note also that $Q_N(\ell,t|\x_0)$ is a monotonously increasing
function of time $t$ by definition.  In the limit $t\to \infty$, one
gets the probability of crossing the threshold $\ell$, i.e., the
probability of stock depletion:
\begin{align}  \nonumber
Q_N(\ell,\infty|\x_0) & = \int\limits_0^\infty dt \, U_N(\ell,t|\x_0)  \\   \label{eq:UN_LaplaceP}
& = 1 - \L^{-1}_{q,\ell} \biggl\{ \frac{[S_q(\infty|\x_0)]^N}{q} \biggr\} \,.
\end{align}
Here, one can distinguish two situations: (i) if any single particle
surely reacts on the partially reactive target $\Gamma$ (i.e.,
$S_q(\infty|\x_0) = 0$), $\ell_t$ will cross any threshold $\ell$ with
probability $Q_N(\ell,\infty|\x_0) = 1$; (ii) in contrast, if the
single particle can survive forever (i.e., $S_q(\infty|\x_0) > 0$) due
to its eventual escape to infinity, then the crossing probability is
strictly less than $1$.  In the latter case, the density
$U_N(\ell,t|\x_0)$ is not normalized to $1$ given that the
first-crossing time can be infinite with a finite probability:
\begin{equation}
\P_{\x_0}\{ \T_{\ell,N} = \infty\} = 1 - Q_N(\ell,\infty|\x_0).  
\end{equation}

The probability density $U_N(\ell,t|\x_0)$ also allows one to compute
the positive integer-order moments of the first-crossing time
(whenever they exist):
\begin{subequations}
\begin{align}
\E_{\x_0} \bigl\{ [\T_{\ell,N}]^k \bigr\} & = \int\limits_0^\infty dt\, t^k \, U_N(\ell,t|\x_0) \\
& = k \int\limits_0^\infty dt\, t^{k-1} \, \bigl(1 - Q_N(\ell,t|\x_0)\bigr)  ,
\end{align}
\end{subequations}
for $k = 1,2,\ldots$, where the second relation is obtained by
integrating by parts under the assumption that $Q_N(\ell,\infty|\x_0)
= 1$ (otherwise the moments would be infinite).  Applying the
inequality (\ref{eq:QN_ineq}), we deduce the monotonous behavior of
all (existing) moments with respect to $\ell$:
\begin{equation}  \label{eq:mean_ineq}
\E_{\x_0}\bigl\{ [\T_{\ell_1,N}]^k \bigr\} \leq \E_{\x_0}\bigl\{ [\T_{\ell_2,N}]^k \bigr\}  \qquad (\ell_1 < \ell_2).
\end{equation}
Expectedly, the moments of the fastest first-passage time $\T_{0,N}$
appear as the lower bounds:
\begin{equation}  \label{eq:mean_bound}
\E_{\x_0}\bigl\{ [\T_{0,N}]^k \bigr\} \leq \E_{\x_0}\bigl\{ [\T_{\ell,N}]^k \bigr\} .
\end{equation}
We stress, however, that the computation and analysis of these moments
is in general rather sophisticated, see an example in
Appendix \ref{sec:mean} for diffusion on the half-line.

\subsection{Short-time behavior}

The short-time behavior of $U_N(\ell,t|\x_0)$ strongly depends on
whether the species are initially released on the stock region or not.
Indeed, if $\x_0 \notin \Gamma$, the species need first to arrive onto
the stock region to initiate its depletion.  Since the survival
probability is very close to $1$ at short times, one can substitute
\begin{equation*}
[S_q(t|\x_0)]^N = \bigl(1 - (1-S_q(t|\x_0))\bigr)^N \approx 1 - N\bigl(1 - S_q(t|\x_0)\bigr) 
\end{equation*}
into Eq. (\ref{eq:UN_def}) to get the short-time behavior
\begin{equation}   \label{eq:UN_t0_x0}
U_N(\ell,t|\x_0) \approx N\, U_1(\ell,t|\x_0)   \qquad (t\to 0).
\end{equation}
As the crossing of any threshold $\ell$ by any species is highly
unlikely at short times, the presence of $N$ independent species
yields an $N$-fold increase of the probability of such a rare event.
%Indeed, one typically has
%$U_1(\ell,t|\x_0) \propto e^{-(\delta + \ell)^2/(4Dt)} \ll 1$, where
%$\delta$ is the distance between the starting point $\x_0$ and the
%stock region (see
%\secUshort).
In fact, 
%To our knowledge, there was no systematic studies of the short-time
%behavior of the probability density $U_1(\ell,t|\x_0)$ for a single
%particle in general confining domains.  
the exact solution (\ref{eq:U1_1d}) for diffusion on the half-line
allows one to conjecture the following short-time asymptotic behavior
in a general domain:
\begin{equation}
U_1(\ell,t|\x_0) \propto t^{-\alpha} \, e^{-(\delta + \ell)^2/(4Dt)}  \qquad (t\to 0),
\end{equation}
where $\delta$ is the distance from the starting point $\x_0$ to the
stock region $\Gamma$, and $\propto$ means proportionality up to a
numerical factor independent of $t$ (as $t\to 0$).  The exponent
$\alpha$ of the power-law prefactor may depend on the domain, even
though we did not observe other values than $\alpha = 3/2$ for basic
examples.  The main qualitative argument in favor of this relation is
that, at short times, any smooth boundary looks as locally flat so
that the behavior of reflected Brownian motion in its vicinity should
be close to that in a half-space, for which the exact solution
(\ref{eq:U1_1d}) is applicable (given that the lateral displacements
of the particle do not affect the boundary local time).  In
particular, one may expect that the geometrical structure of the
domain and of the stock region may affect only the proportionality
coefficient in front of this asymptotic form.  For instance, the exact
solution (\ref{eq:U1_3d}) for diffusion outside a ball of radius $R$
contains the supplementary factor $e^{-\ell/R} R/|\x_0|$, which is not
present in the one-dimensional setting.  Similarly, the short-time
asymptotic relation for $U_1(\ell,t|\x_0)$ in the case of diffusion
outside a disk of radius $R$, that was derived in \cite{Grebenkov21a},
has the factor $e^{-\ell/(2R)} (R/|\x_0|)^{1/2}$.  In both cases, the
additional, non-universal prefactor depends on the starting point
$|\x_0|$ and accounts for the curvature of the boundary via
$e^{-\ell/R}$ or $e^{-\ell/(2R)}$.  Further development of asymptotic
tools for the analysis of the short-time behavior of
$U_1(\ell,t|\x_0)$ in general domains presents an interesting
perspective.

The situation is different when the species are released on the stock
region ($\x_0\in \Gamma$) so that the depletion starts immediately.
The analysis of the short-time behavior is more subtle, while the
effect of $N$ is much stronger.  In Appendix \ref{sec:U_1d}, we
derived the short-time asymptotic formula (\ref{eq:UN_1d_t0}) by using
the explicit form of the survival probability for diffusion on the
half-line with the stock region located at the origin.  This behavior
is valid in the general case because a smooth boundary of the stock
region ``looks'' locally flat at short times.  Moreover, the effect of
local curvature can be partly incorporated by rewriting the
one-dimensional result as
\begin{equation}  \label{eq:UN_short}
U_N(t,\ell|\x_0) \simeq 2^{N-1} \, N \, U_1(Nt,\ell|\x_0)  \quad (t\to 0).
\end{equation}
i.e., the effect of $N$ independent species is equivalent at short
times to an $N$-fold increase of time $t$ for a single species and a
multiplication by a factor $2^{N-1} N$ whose probabilistic origin is
clarified in Appendix \ref{sec:U_1d}.

%%
%\begin{equation}  \label{eq:UN_short_1d}
%U_N(\ell,t|0) \simeq 2^{N-1} \frac{\ell \, e^{-\ell^2/(4NDt)}}{\sqrt{4\pi NDt^3}}  \qquad (t\to 0).
%\end{equation}
%Apart from the combinatorial prefactor $2^{N-1}$ (see discussion in
%\secUd), the right-hand side is the probability density
%of the first-crossing time $\T_{\ell,1}$ for a single particle that
%starts from $x_0 = 0$ and diffuses on the half-line with the $N$-fold
%larger diffusion coefficient.  The latter is identical to the famous
%L\'evy-Smirnov probability density of the first-passage time on the
%half-line.  In a general setting, the species started on the stock
%region does not ``recognize'' its eventual curvature at short times,
%as if the stock region was flat.  As a consequence, the short-time
%behavior of $U_N(t,\ell|\x_0)$ is similar to that in the half-space,
%$\R^{d-1}\times \R_+$, which in turn is identical to
%Eq. (\ref{eq:UN_short_1d}) for the half-line.  The effect of curvature
%can be partly incorporated by using the approximation
%%
%\begin{equation}  \label{eq:UN_short}
%U_N(t,\ell|\x_0) \simeq 2^{N-1} \, N \, U_1(Nt,\ell|\x_0)  \quad (t\to 0),
%\end{equation}
%i.e., the effect of $N$ independent species is equivalent at short
%times to an $N$-fold increase of time $t$ for a single species and a
%multiplication by factor $2^{N-1} N$.

As the cumulative distribution function $Q_N(\ell,t|\x_0)$ is obtained
by integrating $U_N(\ell,t'|\x_0)$ over $t'$ from $0$ to $t$, one can
easily derive its asymptotic behavior from Eqs. (\ref{eq:UN_t0_x0},
\ref{eq:UN_short}):
\begin{align}  \label{eq:QN_short1}
Q_N(\ell,t|\x_0) & \approx N Q_1(\ell,t|\x_0)  \qquad (\x_0 \notin \Gamma), \\    \label{eq:QN_short2}
Q_N(\ell,t|\x_0) & \approx 2^{N-1} \, Q_1(\ell,Nt|\x_0)  \quad (\x_0 \in \Gamma).
\end{align}

\subsection{Long-time behavior}

The long-time behavior of the probability density $U_N(\ell,t|\x_0)$
is related via Eq. (\ref{eq:UN_def}) to that of the survival
probability $S_q(t|\x_0)$, according to which we distinguish four
situations:
\begin{equation}  \label{eq:Sq_asympt}
S_q(t|\x_0) \simeq \left\{ \begin{array}{l l}  e^{-D\lambda_0^{(q)} t} \, \psi_q(\x_0) & (\textrm{class I}), \\
t^{-\alpha} \, \psi_q(\x_0) & (\textrm{class II}), \\
(\ln t)^{-\alpha} \psi_q(\x_0) & (\textrm{class III}), \\
S_q(\infty|\x_0) + t^{-\alpha} \psi_q(\x_0) & (\textrm{class IV}), \\ \end{array}  \right.  
\end{equation}
where $\lambda_0^{(q)}$ is the smallest eigenvalue of the Laplace
operator in $\Omega$ with mixed Robin-Neumann boundary condition
(\ref{eq:Sq_BC}), $\alpha > 0$ is a persistence exponent
\cite{Redner,Bray13,Levernier19}, and $\psi_q(\x_0)$ is a
domain-specific function of $\x_0$ and $q$.  Even though the above
list of asymptotic behaviors is not complete (e.g., there is no
stretched-exponential behavior observed in disordered configurations
of traps \cite{Kayser83,Kayser84}), these classes cover the majority
of cases studied in the literature.  For instance, class I includes
all bounded domains, in which the spectrum of the Laplace operator is
discrete, allowing for a spectral expansion of the survival
probability and yielding its exponentially fast decay as $t\to\infty$.
For unbounded domains, the long-time behavior of $S_q(t|\x_0)$ is less
universal and strongly depends on the space dimensionality $d$ and the
shape of the domain \cite{Redner,Bray13,Levernier19,Guerin21}.  For
instance, class II includes: (a) the half-line or, more generally, a
half-space, with $\alpha = 1/2$ and explicitly known form of
$\psi_q(\x_0)$ (see Appendix \ref{sec:half-line}); (b) a perfectly
reactive wedge of angle $\theta$ in the plane, with $\alpha =
\pi/(2\theta)$ \cite{Redner}; (c) a perfectly reactive cone in three
dimensions, with a nontrivial relation between $\alpha$ and the cone
angle \cite{Redner}.  The exterior of a disk in the plane and the
exterior of a circular cylinder in three dimensions are examples of
domains in class III \cite{Redner,Levitz08,Grebenkov21a}.  Class IV
includes the exterior of a bounded set in three dimensions, in which a
particle can escape to infinity and thus never react on the target,
with the strictly positive probability $S_q(\infty |\x_0)$ (see
Appendix \ref{sec:ball}).

It is easy to check that Eq. (\ref{eq:UN_def}) implies the long-time
behavior:
\begin{equation}  \label{eq:UN_asympt}
U_N(\ell,t|\x_0) \simeq
\left\{ \begin{array}{l l}  N\alpha \, t^{-N\alpha-1} \, \Psi_N(\x_0,\ell) & (\textrm{class II}), \\
\displaystyle \frac{N\alpha \, t^{-1}}{(\ln t)^{N\alpha+1}}\, \Psi_N(\x_0,\ell)  & (\textrm{class III}), \\
N\alpha \, t^{-\alpha-1}\, \Psi_N(\x_0,\ell) & (\textrm{class IV}), \\ \end{array}  \right.  
\end{equation}
where $\Psi_N(\x_0,\ell) = \L_{q,\ell}^{-1} \{ [\psi_q(\x_0)]^N/q\}$
for classes II and III, and $\Psi_N(\x_0,\ell) = \L_{q,\ell}^{-1} \{
[S_q(\infty|\x_0)]^{N-1} \psi_q(\x_0)/q\}$ for class IV.  One also
gets
\begin{align}  \label{eq:QN_asympt}
Q_N(\ell,t|\x_0) & \simeq  Q_N(\ell,\infty|\x_0)  \\  \nonumber
& - \left\{ \begin{array}{l l}  t^{-N\alpha} \, \Psi_N(\x_0,\ell) & (\textrm{class II}), \\
\displaystyle (\ln t)^{-N\alpha}\, \Psi_N(\x_0,\ell)  & (\textrm{class III}), \\
N\, t^{-\alpha}\, \Psi_N(\x_0,\ell) & (\textrm{class IV}), \\ \end{array}  \right.  
\end{align}
where $Q_N(\ell,\infty|\x_0)$ is the crossing probability.  In turn,
the asymptotic behavior in bounded domains (class I) is more subtle
and will be addressed elsewhere (see discussions in
\cite{Grebenkov19b,Grebenkov20,Grebenkov20b,Grebenkov20c} for a single
particle).

According to Eqs. (\ref{eq:UN_asympt}, \ref{eq:QN_asympt}), the effect
of multiple species
%on the long-time behavior
%of the first-crossing time 
strongly depends on the geometric structure of the domain.  For class
II, each added species enhances the power law decrease of the
probability density.  In particular, the mean first-crossing time is
infinite for $N \leq 1/\alpha$ and finite for $N > 1/\alpha$.  For
instance, when the species diffuse on the half-line, the mean
first-crossing time is finite for $N > 2$ and scales as $N^{-2}$ at
large $N$ (see Appendix \ref{sec:mean}).  Higher-order moments are
getting finite as $N$ increases.  This effect is greatly diminished
for class III, in which the ``gain'' from having multiple species is
just in powers of the logarithm of $t$.  As a consequence, the mean
first-crossing time remains infinite for any $N$, despite the
recurrent nature of diffusion when each species returns infinitely
many times to the stock region.  For domains of class IV, the
transient character of diffusion implies that each species may
encounter the stock region a limited number of times before leaving it
forever by escaping to infinity with a finite probability.  As a
consequence, the probability density decays as $t^{-\alpha-1}$ for any
$N$, and the number of species affects only the prefactor in front of
this universal form.
%We conclude that the
%above asymptotic relations help to address two practical questions:
%how long does the stock remain available and how the initial amount of
%resources affects the depletion time?
%
Note that the stock depletion is certain (with probability $1$) for
classes II and III; in turn, this probability is below $1$ for class
IV but it approaches $1$ exponentially rapidly as $N$ increases, see
Eq. (\ref{eq:UN_LaplaceP}).

\subsection{Example of a spherical stock region}

\begin{figure}
\begin{center}
\includegraphics[width=88mm]{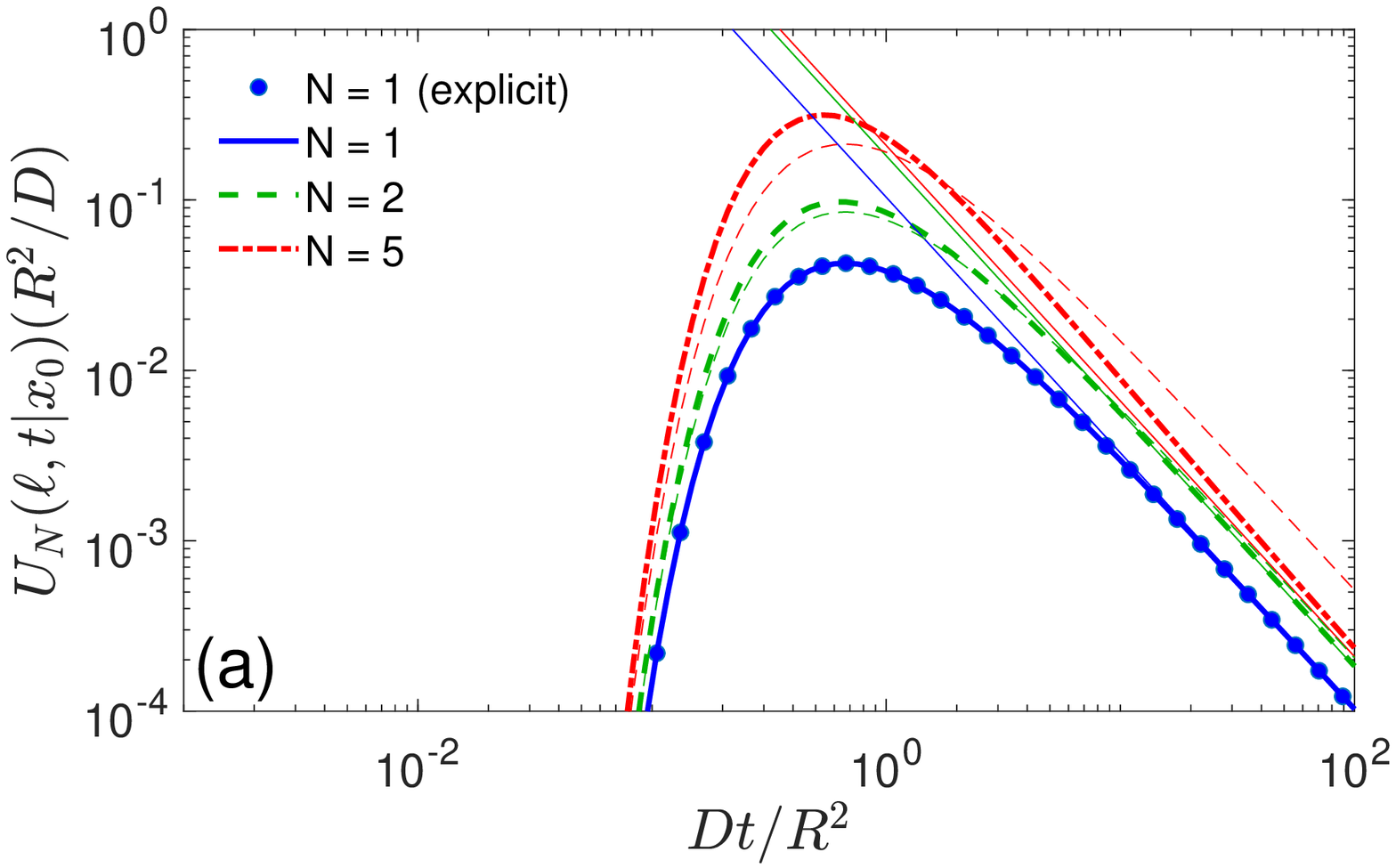} % UN_3d_r2new.eps}
\includegraphics[width=88mm]{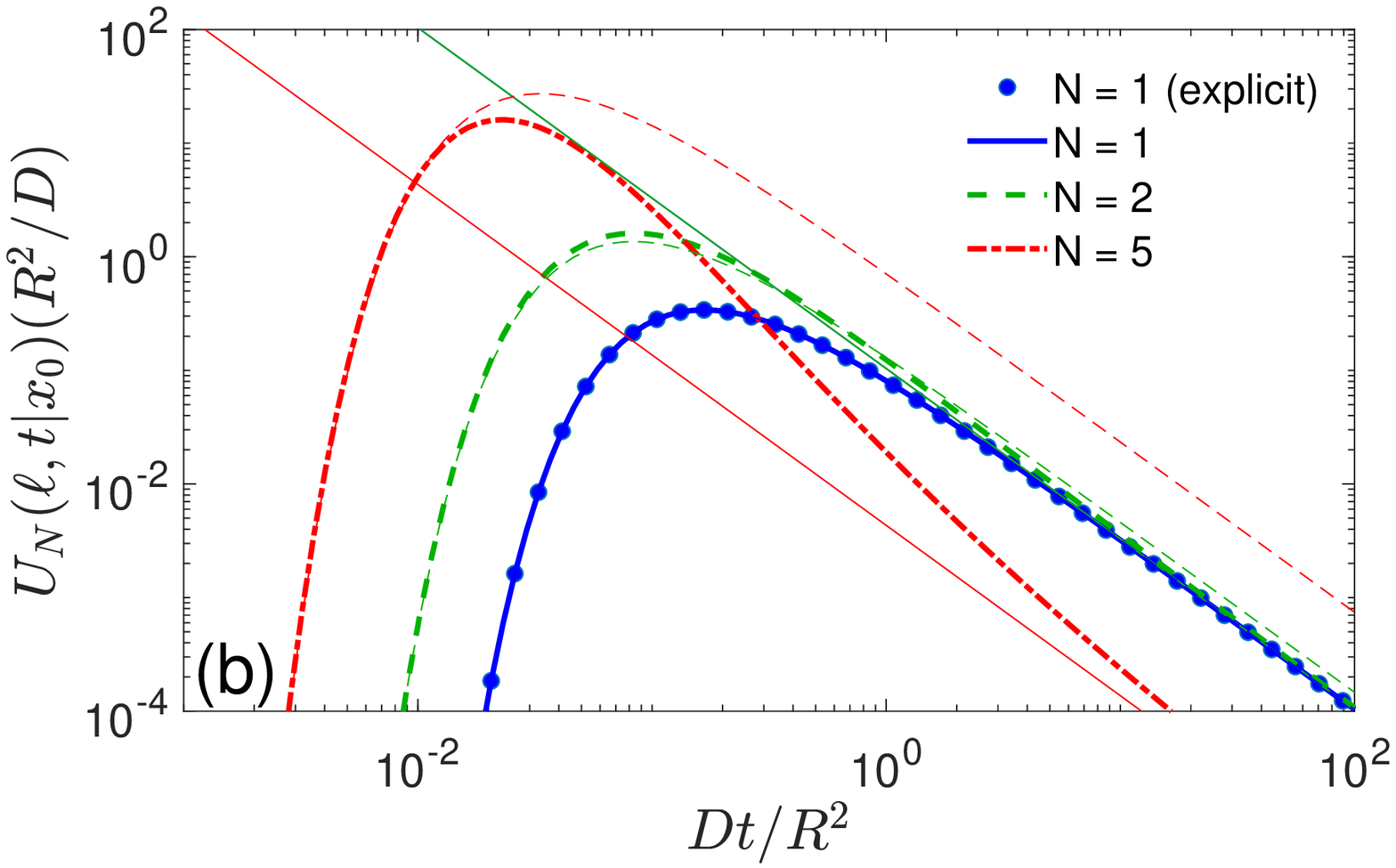} % UN_3d_r1.eps}
\end{center}
\caption{
Probability density function $U_N(\ell,t|\x_0)$ of the first-crossing
time $\T_{\ell,N}$ for $N$ species diffusing in the exterior of a
spherical stock region of radius $\rrho$, with $\ell = \rrho$, and
$|\x_0| = 2\rrho$ {\bf (a)} and $|\x_0| = \rrho$ {\bf (b)}.  Symbols
present the explicit form (\ref{eq:U1_3d}) for a single species,
whereas thick lines show the result of numerical integration in
Eq. (\ref{eq:UN_3d}), see Appendix \ref{sec:numerical}.  Thin solid
lines indicate the long-time asymptotic relation (\ref{eq:UN_asympt}),
with $\alpha = 1/2$ and $\Psi_N(\x_0,\ell)$ is given by
Eq. (\ref{eq:PsiN_3d}); in turn, thin dashed lines present the
short-time behavior in Eq. (\ref{eq:UN_t0_x0}) for panel (a) and
Eq. (\ref{eq:UN_short}) for panel (b).}
\label{fig:UN_3d}
% [U1,U2,U5, t] = A_localtime6_UN_3d_fig2(U1,U2,U5); % r0=2 
% [U1,U2,U5, t] = A_localtime6_UN_3d_fig(U1,U2,U5);  % r0=1
\end{figure}

To illustrate the properties of the first-crossing time $\T_{\ell,N}$,
we consider $N$ species diffusing in the three-dimensional space and
searching for a spherical stock region of radius $\rrho$.  In this
setting (class IV), the survival probability $S_q(t|\x_0)$ has an
exact explicit form that allowed us to compute numerically the
probability density $U_N(\ell,t|\x_0)$,
%according to Eq. (\ref{eq:UN_def}), 
see Appendix \ref{sec:numerical} for details.  For $N = 1$, this
density gets an explicit form \cite{Grebenkov20c}:
\begin{equation} \label{eq:U1_3d}
U_1(\ell,t|\x_0) = \frac{\rrho \, e^{-\ell/\rrho}}{|\x_0|} \, \frac{|\x_0|-\rrho+\ell}{\sqrt{4\pi Dt^3}} e^{-(|\x_0| - \rrho +\ell)^2/(4Dt)} .
\end{equation}
Setting $\ell = 0$, one retrieves the probability density of the
first-passage time for a perfectly absorbing sphere
\cite{Smoluchowski17}.

Figure \ref{fig:UN_3d} shows the probability density
$U_N(\ell,t|\x_0)$ and its asymptotic behavior for a particular
threshold $\ell = \rrho$.  When the species start a distance away from
the stock region (panel (a)), $U_N(\ell,t|r_0)$ looks as being just
``shifted'' upwards by increasing $N$, in agreement with the
short-time behavior in Eq. (\ref{eq:UN_t0_x0}).  In particular, the
most probable first-crossing time remains close to that of a single
species.  Here, the species need first to reach the stock region, so
that speed up of the depletion by having many species is modest.  The
situation is drastically different when the species start on the stock
region (panel (b)).  In this case, some species may stay close to the
stock region, repeatedly returning to it and rapidly consuming its
resources.  One sees that the total boundary local time reaches a
prescribed threshold $\ell$ much faster, and the probability density
$U_N(\ell,t|r_0)$ is shifted towards shorter times as $N$ increases.
In both panels, the short-time and long-time asymptotic relations
derived above are accurate.  We stress that the mean first-crossing
time and higher-order moments are infinite and thus not informative
here.
%One sees that both short-time and long-time
%asymptotic relations derived above are accurate, and they emphasize
%the crucial role of the starting point.  
Some other aspects of this depletion problem, such as the cumulative
distribution function $Q_N(\ell,t|\x_0)$, the probability of
depletion, and their dependence on $N$, are discussed in Appendix
\ref{sec:ball}.  In turn, Appendix \ref{sec:half-line} presents the
study of diffusion on the half-line (class II).

\section{Discussion and Conclusion}
%
%In this letter, we presented the first theoretical study of the stock
%depletion by a population of diffusing species.  We focused on a large
%class of depletion processes when each species consumes a unit of
%resources upon each encounter with the stock region.  As these
%encounters are naturally characterized by the boundary local time, the
%depletion time turns out to be the first-crossing time $\T_{\ell,N}$
%of a prescribed threshold $\ell$ (the initial amount of resources) by
%the total boundary local time of $N$ species.  Relying on a recently
%developed encounter-based approach to diffusion-controlled reactions
%\cite{Grebenkov20}, we managed to express the probability density
%$U_N(\ell,t|\x_0)$ of $\T_{\ell,N}$ in terms of the survival
%probability $S_q(t|\x_0)$ of a single species.  The fundamental
%relation (\ref{eq:UN_def}) allowed us to compute this density in
%different settings and to show the impact of the number of species
%onto the asymptotic behavior of $U_N(\ell,t|\x_0)$ at both short and
%long times.

As depletion of resources is one of the major modern problems,
numerous former studies addressed various aspects of this phenomenon.
For instance, B\'enichou {\it et al.} investigated
depletion-controlled starvation of a diffusing forager and related
foraging strategies
\cite{Benichou14,Chupeau16,Benichou16,Chupeau17,Bhat17,Benichou18}.
These studies focused on the forager itself and on the role of
depletion on its survival.  In contrast, our emphasis was on the
dynamics of stock depletion, i.e., how fast available resources are
exhausted by a population of diffusing species.  To our knowledge,
this problem was not previously addressed, and the present work
settles a first theoretical ground for further explorations of this
important topic in several directions.

(i) While we focused on a fixed starting point $\x_0$ for all species,
an extension of our results to the case of independent randomly
distributed starting points is straightforward.  In particular, the
major difference between Eq. (\ref{eq:UN_t0_x0}) for $\x_0 \notin
\Gamma$ and Eq. (\ref{eq:UN_short}) for $\x_0 \in
\Gamma$ suggests that the form of the initial distribution of
$\x_0$ in the vicinity of the stock region may strongly affect the
short-time behavior of the probability density $U_N(\ell,t|\x_0)$.

(ii) For diffusion in bounded domains, the long-time behavior of the
probability density $U_N(\ell,t|\x_0)$ requires a subtle asymptotic
analysis of the ground eigenmode of the Laplace operator as a function
of the implicit reactivity parameter $q$; the role of the geometric
confinement remains to be elucidated.

(iii) In the considered model of non-renewable resources, the stock
region is depleted upon each encounter with each diffusing species.
This assumption can be relaxed in different ways.  For instance, one
can consider a continuous-time supply of resources, for which the
problem is equivalent to finding the first-crossing time of a
deterministic time-dependent threshold $\ell(t)$.  Alternatively,
replenishment of resources can be realized at random times, as a sort
of stochastic resetting.  If the resetting times are independent from
diffusion of species, one may apply the renewal theory, which was
successful in describing diffusion with resetting
\cite{Evans11,Chechkin18,Evans20}.  Yet another option consists of
implementing a dynamic regeneration of consumed resources on the stock
region (like a natural regeneration of forests).  Finally, one can
also include more sophisticated consumption mechanisms when resources
are distributed to each species depending on the number of its
previous encounters with the stock region (e.g., a species receives
less resources at its next return to the stock region).  This
mechanism and its theoretical implementation resemble the concept of
encounter-dependent reactivity in diffusion-controlled reactions
\cite{Grebenkov20}.

(iv) Another direction consists in elaborating the properties of
species.  First, one can incorporate a finite lifetime of diffusing
species and analyze the stock depletion by ``mortal'' walkers
\cite{Meerson15,Grebenkov17d}.  The effect of diversity of species
(e.g., a distribution of their diffusion coefficients) can also be
analyzed.  Second, dynamics beyond ordinary diffusion can be
investigated; for instance, the distribution of the boundary local
time was recently obtained for diffusion with a gradient drift
\cite{Grebenkov22}.  The knowledge on the survival probability of more
sophisticated stochastic dynamics, such as diffusing diffusivity or
switching diffusion models
\cite{Godec17,Lanoiselee18,Sposini19,Grebenkov19f}, can potentially be
employed in the analysis of the stock depletion problem.  Further
incorporation of interactions between species (such as communications
between ants, bees or birds) may allow to model advanced strategies of
faster stock depletion that are common in nature.  On the other hand,
one can consider multiple stock regions and inquire on their optimal
spatial arrangments or replenishment modes to construct sustainable
supply networks.

The combination of these complementary aspects of the stock depletion
problem will pave a way to understand and control various depletion
phenomena in biology, ecology, economics and social sciences.

\begin{acknowledgments}
The author acknowledges a partial financial support from the Alexander von
Humboldt Foundation through a Bessel Research Award. 
\end{acknowledgments}

%%%%%%%%%%%%%%%%%%%%%%%%%%%%%%%%%%%%%%%%%%%%%%%%%%%%%%%%%%%%%%%%%%%%%%
\appendix

%\subsection{Examples of our classification}
%\label{sec:classification}
%
%Equation (\ref{eq:Sq_asympt}) provides a partial classification of
%confining domains according to the long-time behavior of the survival
%probability $S_q(t|\x_0)$.  Here we give some examples.

\section{Diffusion on a half-line}
\label{sec:half-line}

In this Appendix, we investigate the stock depletion problem by a
population of species diffusing on the half-line, $\Omega =
\R_+$.  We first recall the basic formulas for a single particle and
then proceed with the analysis for $N$ particles.  We stress that this
setting is equivalent to diffusion in the half-space $\R^{d-1} \times
\R_+$ because the boundary local time is not affected by lateral
displacements of the particles along the hyperplane $\R^{d-1}$.

\subsection{Reminder for a single particle}

For the positive half-line with partially reactive endpoint $0$, the
survival probability reads \cite{Redner} 
\begin{equation}  \label{eq:Sq_1d}
S_q(t|x_0) = \erf(z_0) + e^{-z_0^2} \erfcx\bigl(z_0 + q\sqrt{Dt}\bigr) ,
\end{equation}
where $\erfcx(z) = e^{z^2} \erfc(z)$ is the scaled complementary error
function, and $z_0 = x_0/\sqrt{4Dt}$.  One has $S_q(t|x_0) \to 1$ as
$q\to 0$, and
\begin{equation}  \label{eq:Sq_1d_asympt}
S_q(t|x_0) \xrightarrow[q\to\infty]{}  S_\infty(t|x_0) = \erf(z_0) + \frac{1}{\sqrt{\pi Dt}}\, q^{-1} + O(q^{-2}),
\end{equation}
where we used the asymptotic behavior of $\erfcx(z)$.  The probability
density of the first-passage time, $H_q(t|x_0) = -\partial_t
S_q(t|x_0)$, is
\begin{equation}
H_q(t|x_0) = qD e^{-z_0^2} \biggl(\frac{1}{\sqrt{\pi Dt}} - q \, \erfcx\bigl(z_0 + q\sqrt{Dt}\bigr) \biggr).
\end{equation}
Note also that
\begin{equation}  \label{eq:Sq_1d_short}
S_q(t|x_0) \simeq 1 - \frac{2\sqrt{Dt}}{x_0 \sqrt{\pi}}  \, \frac{2qDt}{x_0 + 2q Dt} \, e^{-x_0^2/(4Dt)}  \quad (t\to 0),
\end{equation}
so that the algebraic prefactor in front of $e^{-x_0^2/(4Dt)}$ is
different for perfectly and partially reactive targets.  In the
long-time limit, one gets
\begin{equation}
S_q(t|x_0) \simeq \frac{x_0 + 1/q}{\sqrt{\pi Dt}} + O(t^{-1})  \qquad (t\to\infty),
\end{equation}
i.e., the half-line belongs to class II according to our
classification in Eq. (\ref{eq:Sq_asympt}), with
\begin{equation}  \label{eq:psi_1D}
\alpha = \frac12 \,, \qquad \psi_q(x_0) = \frac{x_0 + 1/q}{\sqrt{\pi D}} \,.
\end{equation}

The probability density of the boundary local time $\ell_t^1$ is
\begin{equation}  \label{eq:rho_1d}
\rho_1(\ell,t|x_0) = \erf\biggl(\frac{x_0}{\sqrt{4Dt}}\biggr) \delta(\ell) + \frac{\exp\bigl(-\frac{(x_0+\ell)^2}{4Dt}\bigr)}{\sqrt{\pi Dt}}  \,,
\end{equation}
while the probability density of the first-crossing time of a
threshold $\ell$ by $\ell_t^1$ reads \cite{Borodin,Grebenkov20c}:
\begin{equation}  \label{eq:U1_1d}
U_1(\ell,t|x_0) = (\ell+x_0) \frac{e^{-(\ell+x_0)^2/(4Dt)}}{\sqrt{4\pi Dt^3}} \,.
\end{equation}
Note that 
\begin{equation}
Q_1(\ell,t|x_0) =
\int\limits_\ell^\infty d\ell' \, \rho_1(\ell',t|x_0) = \erfc\biggl(\frac{x_0+\ell}{\sqrt{4Dt}}\biggr) .
\end{equation}
The most probable first-crossing time corresponding to the maximum of
$U_1(t,\ell|x_0)$ is
\begin{equation}
t_{\rm mp,1} = \frac{(x_0 + \ell)^2}{6D} \,.
\end{equation}

\subsection{PDF of the total boundary local time}
\label{sec:1D_rhoN}

The probability density of the total boundary local time $\ell_t$ is
determined via the inverse Laplace transform in
Eq. (\ref{eq:rhoN_ILq}).  In Appendix \ref{sec:numerical}, we provide
an equivalent representation (\ref{eq:rhoN_ell}) in terms of the
Fourier transform, which is more suitable for the following analysis.
Substituting $S_q(t|x_0)$ from Eq. (\ref{eq:Sq_1d}), we get
\begin{equation}  \label{eq:rhoN_1d}
\rho_N(\ell,t|x_0) = \bigl( \erf(z_0) \bigr)^N \delta(\ell) + \frac{I_N(\ell/\sqrt{Dt}, z_0)}{\sqrt{Dt}}  \,,
\end{equation}  
where
\begin{align}  \label{eq:IN_def}
& I_N(\lambda,z_0) = \int\limits_{-\infty}^\infty \frac{dq}{2\pi} \, e^{iq\lambda} \\  \nonumber
& \quad \times \biggl[\biggl( \erf(z_0) + e^{-z_0^2} \erfcx(z_0 + iq)\biggr)^{N} - \bigl( \erf(z_0) \bigr)^N \biggr] \,.
\end{align}  

The small-$\ell$ asymptotic behavior of this density can be obtained
as follows.  We distinguish two cases: $z_0 > 0$ or $z_0 = 0$.  In the
former case, we find
\begin{equation}  \label{eq:IN_lam0}
I_N(\lambda,z_0) = \frac{N e^{-z_0^2} \bigl( \erf(z_0) \bigr)^{N-1}}{\sqrt{\pi}} + o(1) \qquad (\lambda \to 0),
\end{equation}
and thus Eqs. (\ref{eq:Sq_1d_asympt}, \ref{eq:rhoN_1d}) imply in the
limit $\ell\to 0$:
\begin{equation}  \label{eq:rhoN_1d_small}
\rho_N(\ell,t|x_0) \simeq \bigl( \erf(z_0) \bigr)^N \delta(\ell)  + \frac{N e^{-z_0^2} \bigl( \erf(z_0) \bigr)^{N-1}}{\sqrt{\pi Dt}} + o(1) .
\end{equation}
In turn, for $z_0 = 0$, one has
\begin{equation}  \label{eq:IN_def_lim}
I_N(\lambda,0) = \int\limits_{-\infty}^\infty \frac{dq}{2\pi} \, e^{iq\lambda}  \biggl( \erfcx(iq)\biggr)^{N}  \,.
\end{equation}  
Note that $w(q) = \erfcx(-iq)$ is the Faddeeva function, which admits
the integral representation:
\begin{align}   \label{eq:w1}
w(q) & = \frac{1}{\sqrt{\pi}} \int\limits_0^\infty dz \, e^{-z^2/4 + iqz} \,.
\end{align}
For large $|q|$, the imaginary part of $w(q)$ behaves as $1/(q
\sqrt{\pi})$, while the real part decays much faster, so that
$\erfcx(-iq) \simeq i/(q\sqrt{\pi})$.  Using this asymptotic behavior,
one can show that
\begin{equation}  \label{eq:IN_def_lim2}
I_N(\lambda,0) \simeq \frac{\lambda^{N-1}}{\pi^{N/2} \, (N-1)!}   \qquad (\lambda \to 0),
\end{equation}
from which
\begin{equation}  \label{eq:rhoN_1d_x0_small}
\rho_N(\ell,t|0) \simeq \frac{\bigl(\ell/\sqrt{Dt}\bigr)^{N-1}}{(N-1)! \, \pi^{N/2} \, \sqrt{Dt}}  \qquad (\ell \to 0) .
\end{equation}

The opposite large-$\ell$ limit relies on the asymptotic analysis of
$I_N(\lambda,z_0)$ as $\lambda \to\infty$.  We re-delegate the
mathematical details of this analysis to Appendix \ref{sec:IN} and
present here the final result based on Eq. (\ref{eq:IN_asympt1}):
\begin{align} \nonumber
\rho_N(\ell,t|x_0) & \approx \frac{1}{\sqrt{\pi Dt}}  \sum\limits_{n=1}^N \binom{N}{n} [\erf(z_0)]^{N-n} \\
& \times e^{-(n x_0 + \ell)^2/(4nDt)} \, \frac{2^{n-1}}{\sqrt{n}}   \quad (\ell\to\infty)\,.
\end{align}
If $\ell \gg Nx_0$, the dominant contribution comes from the term with
$n = N$ that simplifies the above expression as:
\begin{equation}   \label{eq:rhoN_1d_large}
\rho_N(\ell,t|x_0) \approx \frac{2^{N-1}}{\sqrt{\pi NDt}}  e^{-(N x_0 + \ell)^2/(4NDt)}  \,.
\end{equation}

\begin{figure}
\begin{center}
\includegraphics[width=88mm]{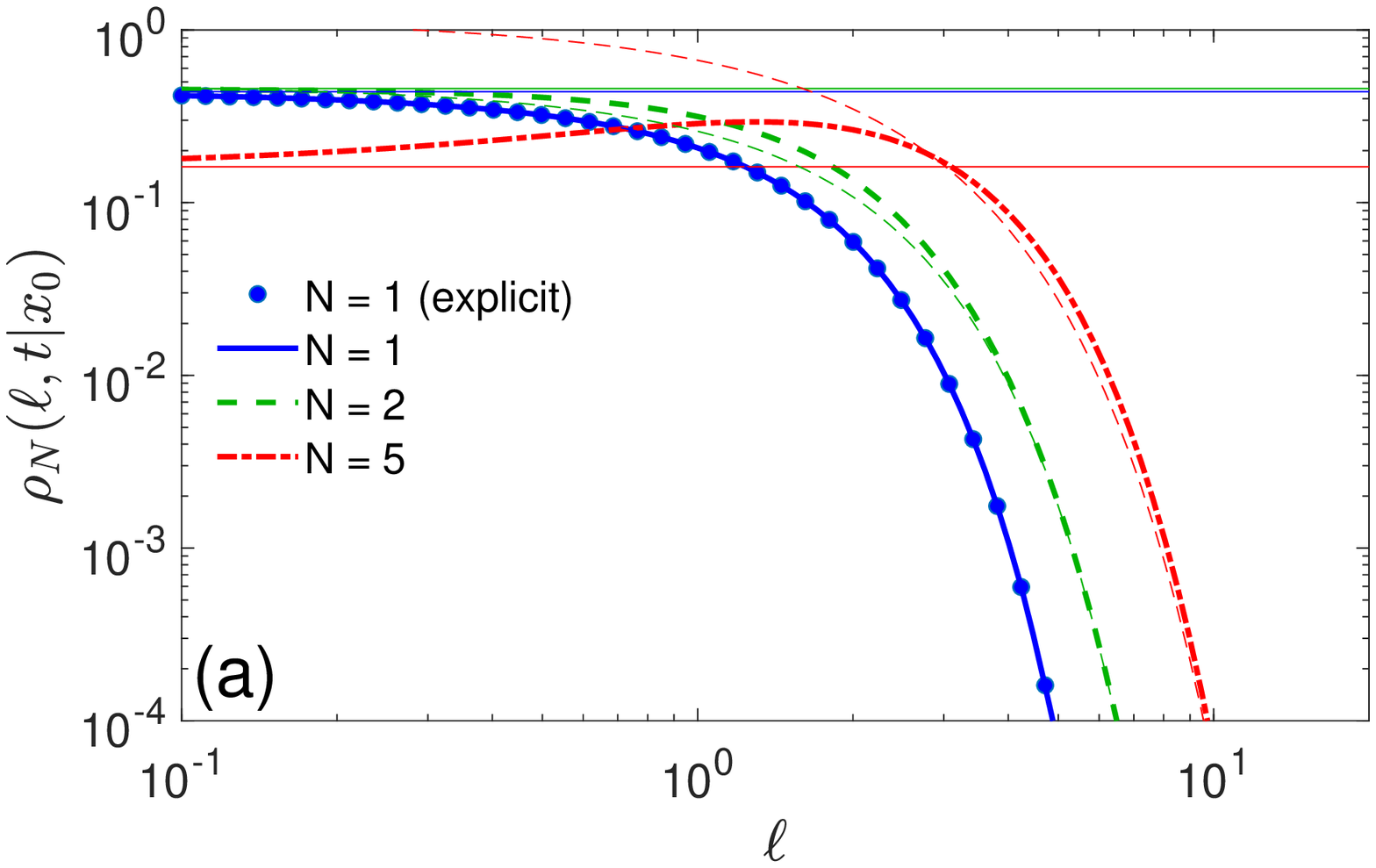} % rhoN_1d_x1.eps}
\includegraphics[width=88mm]{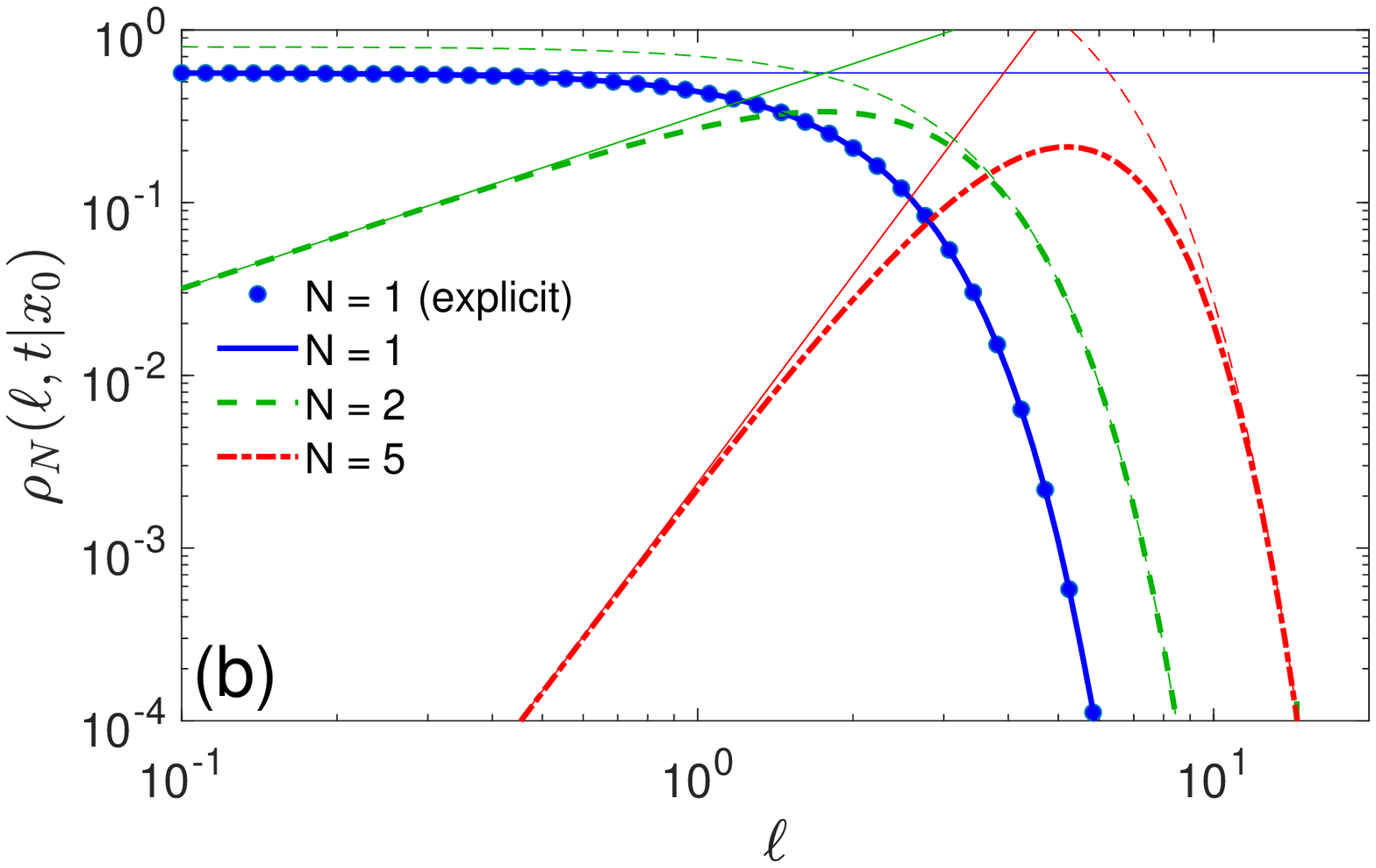} % rhoN_1d_x0.eps}
\end{center}
\caption{
Probability density function $\rho_N(\ell,t|x_0)$ of the total
boundary local time $\ell_t$ for $N$ particles diffusing on the
half-line, with $t = 1$, $D = 1$, and $x_0 = 1$ {\bf (a)} and $x_0 =
0$ {\bf (b)}.  Symbols present the explicit form (\ref{eq:rho_1d}) for
a single particle, whereas thick lines show the numerical integration
in Eqs. (\ref{eq:rhoN_1d}, \ref{eq:IN_def}).  Thin dashed lines
present the large-$\ell$ asymptotic relation (\ref{eq:rhoN_1d_large}),
while thin solid lines indicate the small-$\ell$ asymptotic relation
(\ref{eq:rhoN_1d_small}) for $x_0 = 1$ and (\ref{eq:rhoN_1d_x0_small})
for $x_0 = 0$, respectively.  In panel (a), only the ``regular'' part
is presented, whereas the explicit term with $\delta(\ell)$ is
excluded. }
\label{fig:rhoN_1d_x0}
% [rho1,rho2,rho5, ell] = A_localtime6_rho_fig(rho1,rho2,rho5);  % x0=0
% [rho1,rho2,rho5, ell] = A_localtime6_rho_fig2(rho1,rho2,rho5);
\end{figure}

We emphasize that this result is applicable for any $N$; moreover, for
$N = 1$, this asymptotic formula is actually exact, see
Eq. (\ref{eq:rho_1d}).  This is in contrast with a Gaussian
approximation which was earlier suggested in the long-time limit for
the case of a single particle \cite{Grebenkov07a,Grebenkov19b}.  In
fact, as the particles are independent, the sum of their boundary
local times $\ell_t^i$ can be approximated by a Gaussian variable,
i.e.,
\begin{equation}  \label{eq:rhoN_Gaussian}
\rho_N(\ell,t|x_0) \simeq \frac{\exp\bigl(-\frac{(\ell - N \E_{x_0} \{ \ell_t^1 \})^2}{2 N\var_{x_0}\{\ell_t^1\}}\bigr)}
{\sqrt{2\pi N \var_{x_0}\{\ell_t^1\}}}  \qquad (\ell \to \infty).
\end{equation}
This relation could also be obtained by using the Taylor expansion of
the integrand function in Eq. (\ref{eq:IN_def}) up to the second order
in $q^2$ for the evaluation of its asymptotic behavior.  The mean and
variance of $\ell_t^1$ that appear in Eq. (\ref{eq:rhoN_Gaussian}),
can be found from the explicit relation (\ref{eq:rho_1d}):
\begin{align}
\E_{x_0} \{ \ell_t^1 \} & = \frac{2\sqrt{Dt}}{\sqrt{\pi}} \, e^{-z_0^2} - x_0 \erfc(z_0), \\
\E_{x_0} \{ [\ell_t^1]^2 \} & = (x_0^2 + 2Dt) \erfc(z_0) - \frac{2x_0\sqrt{Dt}}{\sqrt{\pi}} \, e^{-z_0^2} , 
\end{align}
from which the variance follows as 
\begin{equation}
\var_{x_0}\{\ell_t^1\} = \E_{x_0} \{ [\ell_t^1]^2 \} - \bigl(\E_{x_0} \{\ell_t^1 \}\bigr)^2 .
\end{equation}
In particular, one gets for $x_0 = 0$:
\begin{equation}
\E_{0} \{ \ell_t^1 \} = \frac{2}{\sqrt{\pi}} \sqrt{Dt} \,, \quad \var_{0}\{\ell_t^1\} = 2Dt (1 - 2/\pi) .
\end{equation}
% see [L1,L2,vL, t] = A_localtime6_ell_mean_fig();
However, this approximation is applicable either in the large $N$
limit due to the central limit theorem, or in the long-time limit, in
which each $\ell_t^i$ is nearly Gaussian.  In particular, the Gaussian
approximation (\ref{eq:rhoN_Gaussian}) does not capture the
large-$\ell$ behavior shown in Fig. \ref{fig:rhoN_1d_x0}.

Figure \ref{fig:rhoN_1d_x0} illustrates the behavior of the
probability density $\rho_N(\ell,t|x_0)$ for several values of $N$.
First, one sees that both small-$\ell$ and large-$\ell$ asymptotic
relations are accurate.  When the particles start away from the stock
region (panel (a)), the regular part of $\rho_N(\ell,t|x_0)$
approaches a constant level, which decreases with $N$ according to
Eq. (\ref{eq:rhoN_1d_small}).  In turn, the effect of multiple
particles onto the small-$\ell$ behavior is much stronger when the
particles are released on the stock region (panel (b)).

\subsection{PDF of the first-crossing time}
\label{sec:U_1d}

Substituting $S_q(t|x_0)$ from Eq. (\ref{eq:Sq_1d}) into the Fourier
representation (\ref{eq:UN_Fourier}) of $U_N(\ell,t|x_0)$, we get
\begin{align*}  \nonumber
U_N(\ell,t|x_0) & = \frac{N \, e^{-z_0^2} }{t} \int\limits_{-\infty}^\infty \frac{dq}{2\pi} \, e^{iq\ell/\sqrt{Dt}} \\  \nonumber
& \times \biggl( \erf(z_0) + e^{-z_0^2} \erfcx(z_0 + iq)\biggr)^{N-1} \\
& \times \biggl(\frac{1}{\sqrt{\pi}} - iq \, \erfcx(z_0 + iq)\biggr).
\end{align*}
Evaluating the derivative of the function $\erfcx(z)$, one can
represent this expression as
\begin{align}  \nonumber
U_N(\ell,t|x_0) & = \frac{1}{t} \biggl[ \biggl(\frac{\ell}{\sqrt{4Dt}} + N z_0\biggr) I_N\bigl(\ell/\sqrt{Dt},z_0\bigr) \\  \label{eq:UN_1d}
& - Nz_0 \erf(z_0) I_{N-1}\bigl(\ell/\sqrt{Dt},z_0\bigr) \biggr],
\end{align}
where $I_N(\lambda,z_0)$ is given by Eq. (\ref{eq:IN_def}).
According to Eq. (\ref{eq:rhoN_1d}), we can also write
\begin{align}   \nonumber
U_N(\ell,t|x_0) & = \frac{1}{2t} \biggl( (\ell + N x_0) \rho_N(\ell,t|x_0) \\
& - Nx_0\, \erf(z_0)\, \rho_{N-1}(\ell,t|x_0) \biggr).
\end{align}
In the particular case $x_0 = 0$, one gets a simpler relation
\begin{align}  
U_N(\ell,t|0) & = \frac{\ell}{2t} \rho_N(\ell,t|0).
\end{align}

The long-time asymptotic behavior of $U_N(\ell,t|x_0)$ is determined
by the first line in Eq. (\ref{eq:UN_asympt}).  Substituting $\alpha$
and $\psi_q(x_0)$ from Eq. (\ref{eq:psi_1D}), we find
\begin{equation}  \label{eq:UN_1d_tinf2}
U_N(\ell,t|x_0) \simeq \frac{N \bigl(x_0/\sqrt{\pi Dt}\bigr)^N}{2t} \sum\limits_{n=0}^N \binom{N}{n} \frac{(\ell/x_0)^n}{n!}   \,. 
\end{equation}
In the limit $x_0 \to 0$, only the term with $n = N$ survives,
yielding as $t\to\infty$
\begin{equation}  \label{eq:UN_1d_tinf}
U_N(\ell,t|0) \simeq  \frac{D/\ell^2}{2\pi^{N/2}(N-1)!}\,  (Dt/\ell^2)^{-1-N/2} .
\end{equation}
As a consequence, the mean first-crossing time is infinite for $N = 1$
and $N=2$, but finite for $N > 2$ (see Appendix \ref{sec:mean} for
details).  For $N = 1$, one retrieves the typical $t^{-3/2}$ decay of
the L\'evy-Smirnov probability density of a first-passage time, see
Eq. (\ref{eq:U1_1d}).

To get the short-time behavior, we treat separately the cases $x_0 >
0$ and $x_0 = 0$.  In the former case, Eqs. (\ref{eq:UN_t0_x0},
\ref{eq:U1_1d}) imply
\begin{equation}   \label{eq:UN_1d_t0_x0}
U_N(t,\ell|x_0) \approx  N (\ell+x_0) \frac{e^{-(\ell+x_0)^2/(4Dt)}}{\sqrt{4\pi Dt^3}}  \qquad (t\to 0),
\end{equation}

The analysis is more subtle for $x_0 = 0$, for which
Eq. (\ref{eq:UN_1d}) is reduced to
\begin{equation}   \label{eq:UN_1d_asympt0}
U_N(\ell,t|0) = \frac{\ell}{2t\sqrt{Dt}} \, I_N\bigl(\ell/\sqrt{Dt},0\bigr) .
\end{equation}
Using the asymptotic relation (\ref{eq:IN_asympt3}), we get the
short-time behavior:
\begin{equation}  \label{eq:UN_1d_t0}
U_N(\ell,t|0) \simeq 2^{N-1}\, \frac{\ell}{\sqrt{4\pi NDt^3}}  \, e^{-\ell^2/(4NDt)} \quad (t\to 0).
\end{equation}
This asymptotic relation coincides with the exact Eq. (\ref{eq:U1_1d})
for $N = 1$.  More generally, the short-time behavior for $N$
particles is given, up to a multificative factor $2^{N-1}$, by the
probability density $U_1(\ell,t|0)$ for a single particle but with an
$N$-fold increase of the diffusion coefficient.

\begin{figure}
\begin{center}
\includegraphics[width=60mm]{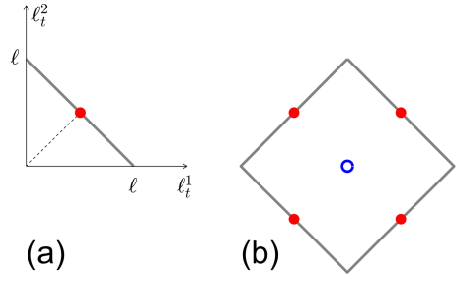} % square.eps}
\end{center}
\caption{
{\bf (a)} Schematic illustration of crossing of a threshold $\ell$ by
$\ell_t = \ell_t^1 + \ell_t^2$ for two particles that corresponds to
crossing the gray line.  Red filled circle indicates the closest
point, through which the crossing is most probable at short times.
{\bf (b)} An equivalent view onto this problem in terms of
two-dimensional Brownian motion that starts from the origin (blue
circle) and searches to exit the rotated square.  Four red filled
circles indicate the closest points through which the exit is most
probable at short times.}
\label{fig:exit}
% A_localtime6_square_exit;
\end{figure}

How can one interpret the prefactor $2^{N-1}$?  For a single particle,
Eq. (\ref{eq:U1_1d}) implies that $U_1(\ell,t|0) =
\frac{\ell}{\sqrt{4\pi Dt^3}} e^{-\ell^2/(4Dt)}$ is identical with the
probability density of the first-passage time to the origin of the
half-line for a particle started a distance $\ell$ away.  In other
words, the threshold $\ell$ effectively increases the distance from
the origin for diffusion on the half-line (see
Refs. \cite{Sapoval05,Grebenkov15} for further discussions on the
geometric interpretation of the boundary local time).  This follows
from the classical fact that the probability law of the boundary local
time in this setting is identical to the probability law of the
reflected Brownian motion $|W_t|$ started from the origin \cite{Levy}.
The reflection symmetry implies that $2 U_1(\ell,t|0)$ also describes
the short-time behavior of the probability density of the first-exit
time from the center of the interval $(-\ell,\ell)$.  Here, the factor
$2$ accounts for the twofold increased probability of the exit event
through two equally distant endpoints.  This interpretation can be
carried on for two particles: the boundary local times $\ell_t^1$ and
$\ell_t^2$ obey the same probability law as two independent reflected
Brownian motions.  As a consequence, the first-crossing of a threshold
$\ell$ by the total boundary local time $\ell_t = \ell_t^1 + \ell_t^2$
is equivalent to the exit from the square of diameter $2\ell$, rotated
by $45^\circ$ (Fig. \ref{fig:exit}).  At short times, the exit is most
probable through vicinities of 4 points that are the closest to the
origin.  As a consequence, $U_2(\ell,t|0) \approx \tfrac12 4
\frac{\ell_2}{\sqrt{4\pi Dt^3}} e^{-\ell_2^2/(4Dt)}$, where $\ell_2 =
\ell/\sqrt{2}$ is the distance from the origin to the edges.  For $N$
particles, the closest distance $\ell_N = \ell/\sqrt{N}$, whereas
there are $2^N$ facets of the hypercube, yielding
Eq. (\ref{eq:UN_1d_t0}).  Even though the exact analogy between the
boundary local time and reflected Brownian motion does not carry on
beyond the half-line, the short-time asymptotic relation is expected
to hold, as illustrated below.

Figure \ref{fig:UN_1d_x0} shows the probability density
$U_N(\ell,t|x_0)$ for several values of $N$.  As expected, the right
(long-time) tail of this density becomes steeper as $N$ increases,
whereas its maximum is shifted to the left (to smaller times).  One
sees that both short-time and long-time relations correctly capture
the asymptotic behavior of $U_N(\ell,t|x_0)$.  At short times, the
starting point $x_0$ considerably affects the probability density.  In
fact, when $x_0 > 0$, the short-time behavior is controlled by the
arrival of any particle to the stock region, and the presence of $N$
particles simply ``shifts'' the density upwards, via multiplication by
$N$ in Eq. (\ref{eq:UN_1d_t0_x0}).  In turn, if the particles start on
the stock region ($x_0 = 0$), the number $N$ significantly affects the
left tail of the probability density, implying a much faster depletion
of resources by multiple particles.

\begin{figure}
\begin{center}
\includegraphics[width=88mm]{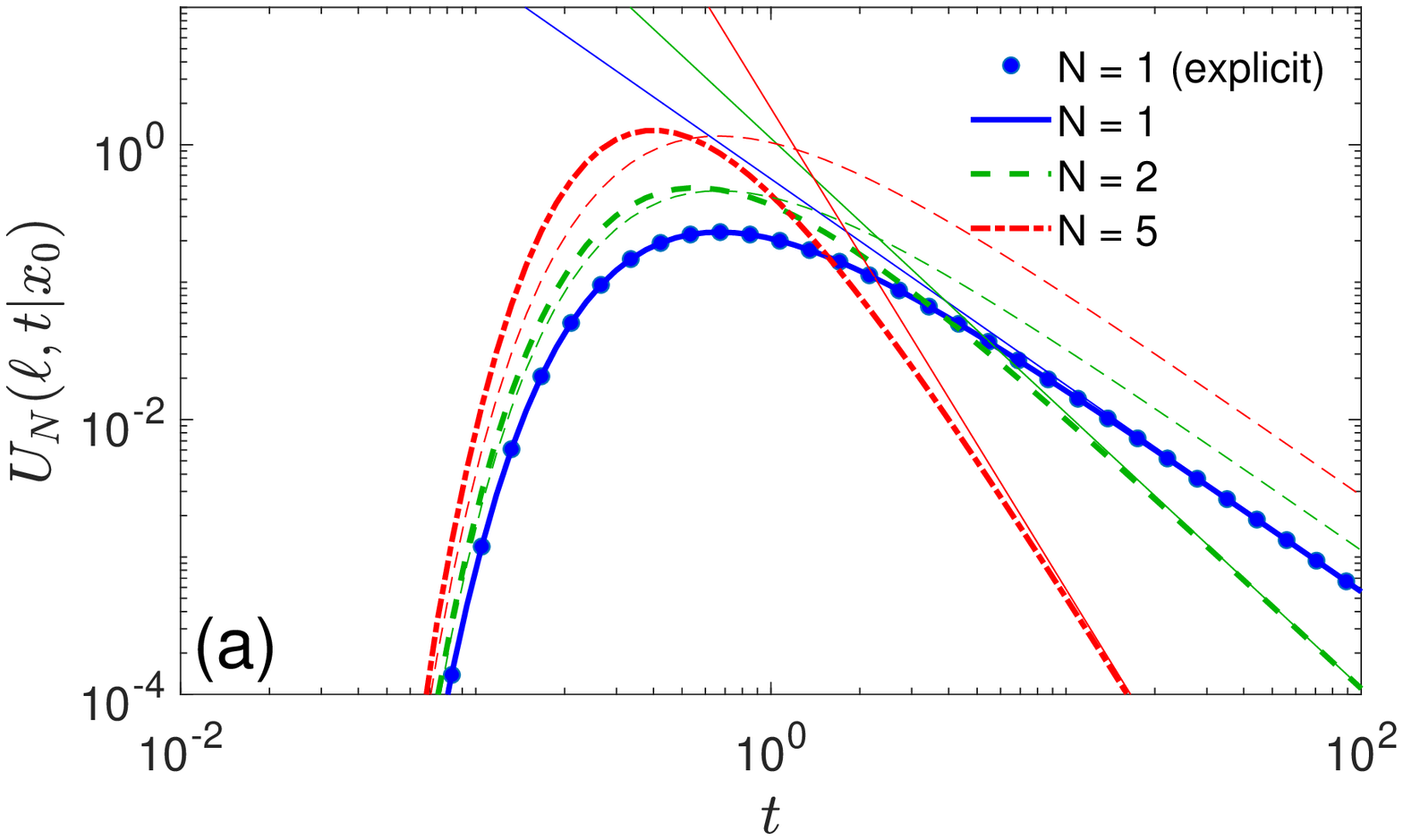} % UN_1d_x1.eps}
\includegraphics[width=88mm]{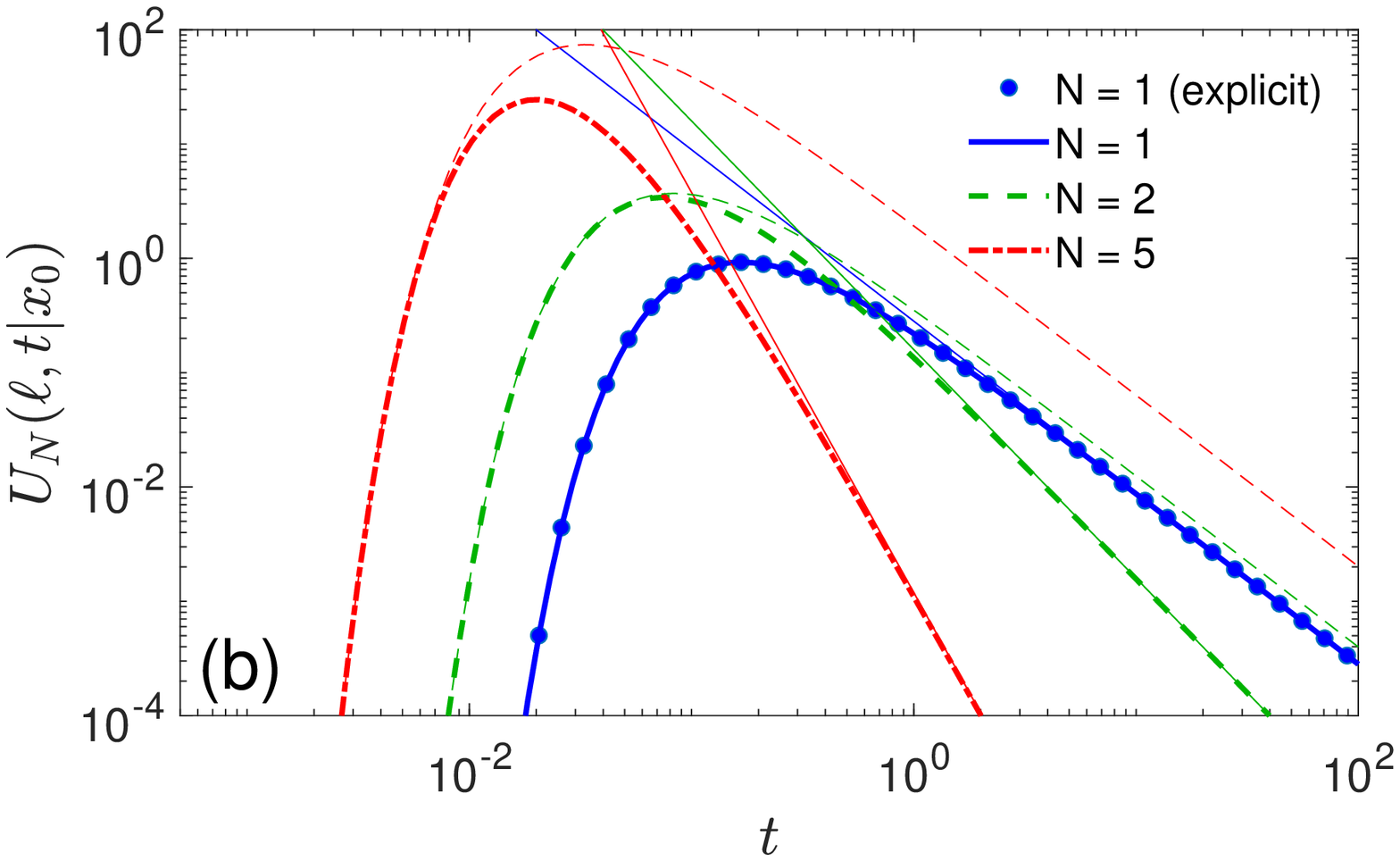} % UN_1d_x0.eps}
\end{center}
\caption{
Probability density function $U_N(\ell,t|x_0)$ of the first-crossing
time $\T_{\ell,N}$ for $N$ particles diffusing on the half-line, with
$\ell = 1$, $D = 1$, and $x_0 = 1$ {\bf (a)} and $x_0 = 0$ {\bf (b)}.
Symbols present the explicit form (\ref{eq:U1_1d}) for a single
particle, whereas thick lines show the numerical integration in
Eqs. (\ref{eq:UN_1d}, \ref{eq:IN_def}).  Thin solid lines indicate the
long-time asymptotic relation (\ref{eq:UN_1d_tinf2}) for $x_0 = 1$ and
(\ref{eq:UN_1d_tinf}) for $x_0 = 0$, respectively.  Thin dashed lines
present the short-time asymptotic relation (\ref{eq:UN_1d_t0}) for
$x_0 = 0$ and (\ref{eq:UN_1d_t0_x0}) for $x_0 = 1$, respectively.}
\label{fig:UN_1d_x0}
% [U1,U2,U5, t] = A_localtime6_UN_fig(U1,U2,U5);  % x0=0
% [U1,U2,U5, t] = A_localtime6_UN_fig2(U1,U2,U5);
\end{figure}

\subsection{Mean first-crossing time}
\label{sec:mean}

Using Eq. (\ref{eq:UN_1d}), one writes the mean first-crossing time as
(whenever it exists)
\begin{align} \nonumber 
\E_{x_0}\{ \T_{\ell,N} \} & = \frac{\ell^2}{D} \int\limits_0^\infty \frac{dy}{y^3} \biggl\{ (y + 2N\xi) I_N(y, y\xi) \\  \label{eq:tau_mean_x0_general}
& - 2N \xi \, \erf(y\xi) I_{N-1}(y, y \xi) \biggr\} ,
\end{align}
with $\xi = x_0/(2\ell)$.  Curiously, the expression for the mean
first-crossing time is more complicated than that for the probability
density.  Since the function $I_N(\lambda,z_0)$ is expressed as an
integral involving the error function, the analysis of this expression
is rather sophisticated.  For this reason, we focus on the particular
case $x_0 = 0$, for which the above expression is reduced to
\begin{equation}
\E_{0}\{ \T_{\ell,N} \} = \frac{\ell^2}{D} \int\limits_0^\infty \frac{dy}{y^2} \, 
\int\limits_{-\infty}^\infty \frac{dq}{2\pi} \, e^{-iqy}  \biggl( \erfcx(- iq)\biggr)^{N}  \,.
\end{equation}
A straightforward exchange of two integrals is not applicable as the
integral of $e^{-iqy}/y^2$ over $y$ diverges.  To overcome this
limitation, we regularize this expression by replacing the lower
integral limit by $\ve$ and then evaluating the limit $\ve \to 0$:
\begin{equation}  \label{eq:tau_mean_x0}
\E_0\{ \T_{\ell,N} \} = \lim\limits_{\ve \to 0} \frac{\ell^2}{D} \int\limits_{-\infty}^\infty \frac{dq}{2\pi} 
\bigl(\erfcx(-iq)\bigr)^N \, F_\ve(q) ,
\end{equation}
where 
\begin{equation}
F_\ve(q) = \frac{e^{-iq\ve}}{\ve} - iq\, \Ei(1,iq\ve) ,
\end{equation}
with $\Ei(1,z)$ being the exponential integral.  The small-$\ve$
expansion of this function reads
\begin{equation}  \label{eq:auxil1}
F_\ve(q) = \ve^{-1} - iq(1-\gamma-\ln(\ve)) + iq\ln(iq) + O(\ve).
\end{equation}
To get a convergent limit in Eq. (\ref{eq:tau_mean_x0}), one has to
show that the integral over $q$ involving the first two terms of this
expansion vanishes, i.e., $J_N^{(0)} = J_N^{(1)} = 0$, where
\begin{equation}
J_N^{(k)} = \pi^{\frac{N}{2}} \int\limits_{-\infty}^\infty \frac{dq}{2\pi} \, q^k  \bigl(\erfcx(-iq)\bigr)^N .
\end{equation}
Let us first consider the integral $J_N^{(0)}$.  Using the
representation (\ref{eq:w1}), we can write
\begin{align*}  \nonumber
J_N^{(0)} & = \int\limits_{-\infty}^\infty \frac{dq}{2\pi} \, 
\int\limits_{\R^N_+} dz_1 \ldots dz_N \, e^{-\frac14(z_1^2+\ldots+z_N^2) + iq(z_1+\ldots+z_N)} \\
& = \int\limits_{\R^N_+} dz_1 \ldots dz_N \, e^{-\frac14(z_1^2+\ldots+z_N^2)} \, \delta(z_1+\ldots+z_N) . 
\end{align*}
For $N = 1$, this integral yields $J_1^{(0)} = \tfrac{1}{2}$, whereas
it vanishes for any $N > 1$.  Similarly, the evaluation of the
integral $J_N^{(1)}$ involves the derivative of the Dirac distribution
and yields $J_2^{(1)} = i/2$, while $J_N^{(1)} = 0$ for any $N > 2$.
We conclude that the limit in Eq. (\ref{eq:tau_mean_x0}) diverges for
$N = 1$ and $N=2$, in agreement with the long-time asymptotic behavior
(\ref{eq:UN_1d_tinf}) of the probability density $U_N(\ell,t|x_0)$.
In turn, for $N > 2$, the limit is finite and is determined by the
integral with the third term in the expansion (\ref{eq:auxil1}):
\begin{equation}  \label{eq:tau_mean_x0_bis}
\E_0\{ \T_{\ell,N} \} = \frac{\ell^2}{D} \int\limits_{-\infty}^\infty \frac{dq}{2\pi}  \, iq\ln(iq) \,
\bigl(\erfcx(-iq)\bigr)^N .
\end{equation}

\begin{figure}
\begin{center}
\includegraphics[width=88mm]{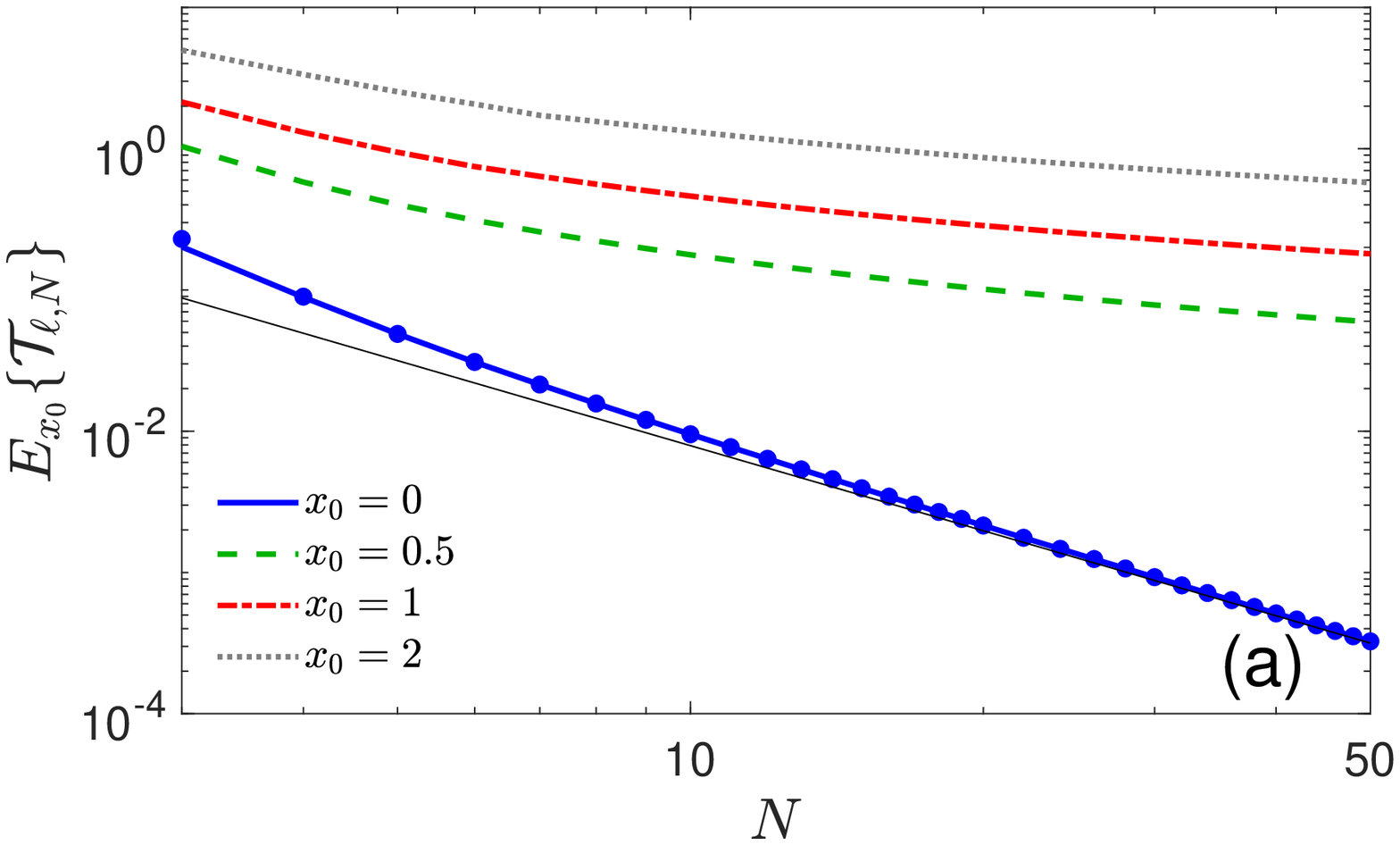} % Tmean_1d.eps}
\includegraphics[width=88mm]{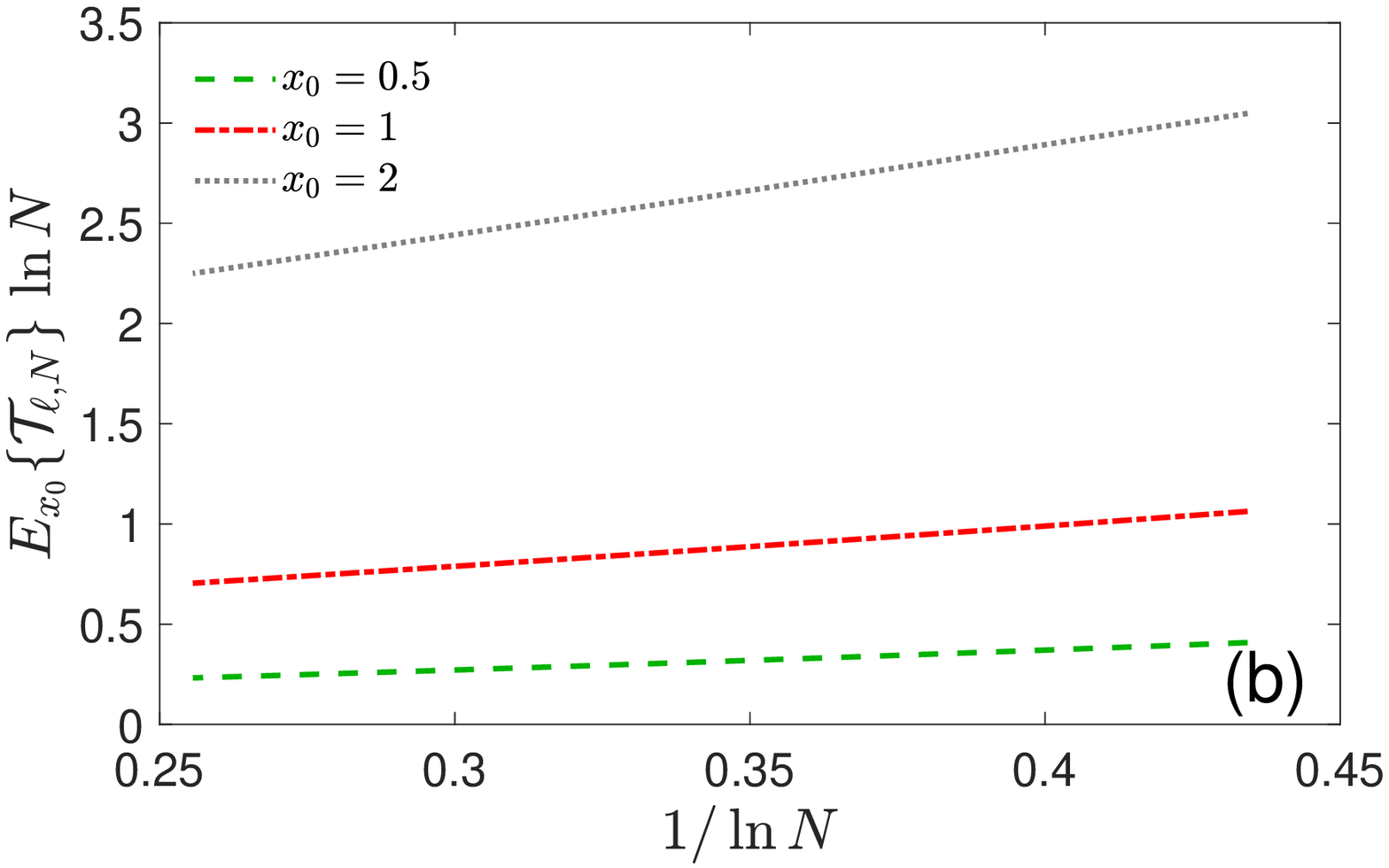} % Tmean_1d_log.eps}
\end{center}
\caption{
{\bf (a)} Mean first-crossing time $\E_{x_0}\{\T_{\ell,N}\}$ as a
function of the number $N$ of particles diffusing on the half-line,
with $\ell = 1$, $D = 1$, and several values of $x_0$ as indicated in
the legend.  Lines show the result of numerical integration in
Eq. (\ref{eq:tau_mean_x0_general}), with the probability density
$U_N(\ell,t|x_0)$ given by Eqs. (\ref{eq:UN_1d}, \ref{eq:IN_def}).
Symbols present the result of numerical integration in
Eq. (\ref{eq:tau_mean_x0_bis2}) for the case $x_0 = 0$.  Thin black
line indicates the asymptotic behavior (\ref{eq:mean_x0_N}).  {\bf
(b)} Another representation of the mean first-crossing time
$\E_{x_0}\{\T_{\ell,N}\}$, multiplied by $\ln N$ and shown as a
function of $1/\ln N$, for $x_0 > 0$.  }
\label{fig:Tmean_1d}
% A_localtime6_tau_mean_fig(T0,T1,T2,T3);
% [T1,T2,T3, N] = A_localtime6_tau_mean_fig2;
\end{figure}

To derive the asymptotic behavior of this integral at large $N$, we
use the Taylor expansion for $\ln(w(q)) \approx iq\frac{2}{\sqrt{\pi}}
- q^2(1-2/\pi) + O(q^3)$ and then approximate the mean as
\begin{equation}  \label{eq:tau_mean_x0_bis2}
\E_0\{ \T_{\ell,N} \} \approx \frac{\ell^2}{D} \, \frac{\pi}{4N^2} \, I_N ,
\end{equation}
with
\begin{equation}
I_N = \int\limits_{-\infty}^\infty \frac{dx}{2\pi}  \, ix\ln(ix\sqrt{\pi}/(2N)) \, e^{ix} \, e^{- x^2/(2z^2)} \,,
\end{equation}
where we rescaled the integration variable as $x = q N(2/\sqrt{\pi})$
and set $z = \sqrt{2N/(\pi-2)}$.  As
\begin{equation*}
\int\limits_{-\infty}^\infty \frac{dx}{2\pi}  \, x \, e^{ix} \, e^{- x^2/(2z^2)} \propto e^{-z^2/2} 
\end{equation*}
is exponentially small for large $N$, one can eliminate the
contribution from a numerical constant under the logarithm that allows
one to write
\begin{equation}
I_N \approx - \int\limits_{0}^\infty \frac{dx}{\pi}  \, x \, \biggl(\frac{\pi}{2} \cos(x) + \sin(x) \, \ln(x)\biggr) \, e^{- x^2/(2z^2)} \,.
\end{equation}
The first term can be evaluated explicitly and yields $1/2$ as $N\to
\infty$.  To proceed with the second term, we employ the
representation $\ln(x) = \lim\limits_{\ve\to 0} (x^\ve - 1)/\ve$ and
exchange the order of integral and limit:
\begin{align*}
I_N & \approx \frac12 - \lim\limits_{\ve\to 0} \frac{1}{\ve} \int\limits_{0}^\infty \frac{dx}{\pi}  \, x^{1+\ve} \, \sin(x) \,  e^{- x^2/(2z^2)} \\
& = \frac12 + \lim\limits_{\ve\to 0} \frac{1}{\ve} \, \frac{\sqrt{2} z^{2+\ve} e^{-z^2/4} 
\bigl(D_{1+\ve}(-z) - D_{1+\ve}(z)\bigr)}{4\sqrt{\pi} \cos(\pi\ve/2)} \,,
\end{align*}
where $D_\nu(z)$ is the Whittaker's parabolic cylinder function, and
we neglected the contribution from $-1/\ve$, which is exponentially
small for large $N$.  For large $z$, $D_{1+\ve}(z)$ is exponentially
small, whereas $D_{1+\ve}(-z)$ behaves as
\begin{equation*}
D_{1+\ve}(-z) \approx - \frac{\sqrt{2\pi}}{\Gamma(-1-\ve)} e^{-i\pi(1+\ve)} z^{-2-\ve} e^{z^2/4}   .
\end{equation*}
As a consequence, one gets
\begin{align}
I_N & \approx \frac12 + \lim\limits_{\ve\to 0} \frac{1}{\ve} \, \frac{e^{-i\pi\ve}}{2 \cos(\pi\ve/2) \Gamma(-1-\ve)} = 1.
\end{align}
We conclude that
\begin{equation} \label{eq:mean_x0_N}
\E_0\{ \T_{\ell,N} \} \approx \frac{\ell^2}{D} \, \frac{\pi}{4} \, N^{-2} \qquad (N\gg 1).
\end{equation}
While the above derivation is not a mathematical proof, it captures
correctly the leading-order behavior of the mean first-crossing time,
see Fig. \ref{fig:Tmean_1d}(a).  A more rigorous derivation and the
analysis of the next-order terms present an interesting perspective.
%
% see \verb|Tmean.mw|] !!!
%{\clr I can even speculate that the next-order term can be well
%approximated as $1/N^3/\sqrt{\ln N}$...}

Equation (\ref{eq:mean_x0_N}) is a rather counter-intuitive result: in
fact, one might expect that the ``speed up'' in crossing the threshold
$\ell$ would be proportional to $N$, i.e., the mean time would be
inversely proportional to $N$.  A similar speed up by $N^2$ was
observed for the mean first-passage time to a perfectly absorbing
target by a population of particles with uniformly distributed initial
positions \cite{Grebenkov20d,Madrid20}.

For the case $x_0 > 0$, one can expect even more sophisticated
behavior.  Indeed, as $\T_{0,N}$ is the fastest first-passage time,
its mean scales with the logarithm of $N$
\cite{Weiss83,Basnayake19,Lawley20,Lawley20b,Lawley20c}
\begin{equation}  \label{eq:Weiss}
\E_{x_0}\{ \T_{0,N} \} \propto \frac{x_0^2}{4D \ln N} \qquad (N\gg 1), 
\end{equation}
i.e., it exhibits a very slow decay with $N$.  For any threshold $\ell
> 0$, the first-crossing time for a single particle naturally splits
into two independent parts: the first-passage time from $x_0$ to the
target, $\T_{0,1}$, and then the first-crossing time $\T_{\ell,1}^{0}$
for a particle started from the target.  The situation is much more
complicated for $N$ particles.  Intuitively, one might argue that it
is enough for a single particle to reach the target and to remain near
the target long enough to ensure the crossing of the threshold $\ell$
by the total boundary local time $\ell_t$, even if all other particles
have not reached the target.  In other words, a single particle may do
the job for the others (e.g., if $\ell_t = \ell_t^1$ and $\ell_t^i =
0$ for all $i=2,3,\ldots,N$).  However, this is not the typical
situation that would provide the major contribution to the mean
first-crossing time.  Indeed, according to the lower bound
(\ref{eq:mean_bound}), the mean first-crossing time $\E_{x_0}\{
\T_{\ell,N} \}$ cannot decrease with $N$ faster than $\E_{x_0}\{
\T_{0,N} \}$, suggesting at least a logarithmically slow decay.

This behavior is confirmed by Fig. \ref{fig:Tmean_1d}(a) showing the
mean first-crossing time $\E_{x_0}\{ \T_{\ell,N} \}$ as a function of
$N$ for a fixed value of $\ell$ and several values of the starting
point $x_0$.  When $x_0 = 0$, we observe the earlier discussed power
law decay (\ref{eq:mean_x0_N}).  In turn, the decay with $N$ is much
slower for $x_0 > 0$.  Multiplying $\E_{x_0}\{ \T_{\ell,N} \}$ by $\ln
N$ and plotting it as a function of $1/\ln N$
(Fig. \ref{fig:Tmean_1d}(b)), we confirm numerically the leading-order
logarithmic behavior (\ref{eq:Weiss}) but with significant
corrections.

\subsection{Large-$\lambda$ asymptotic analysis}
\label{sec:IN}

In this section, we present the details of the large-$\lambda$
asymptotic analysis of the function $I_N(\lambda,z_0)$ defined by
Eq. (\ref{eq:IN_def}).  Using the binomial expansion, one gets
\begin{equation}  \label{eq:IN_z01}
I_N(\lambda,z_0) = \sum\limits_{n=1}^N \binom{N}{n} [\erf(z_0)]^{N-n} e^{-nz_0^2} \, i_n(\lambda,z_0),
\end{equation}
where
\begin{equation}
i_n(\lambda,z_0) = \int\limits_{-\infty}^\infty \frac{dq}{2\pi} \, e^{iq\lambda} \bigl[w(iz_0-q)\bigr]^{n}  ,
\end{equation}  
and we used the Faddeeva function $w(z)$ to express $\erfcx(z_0 +
iq)$.  To evaluate the large-$\lambda$ asymptotic behavior of the
integral $i_n(\lambda,z_0)$, we employ the integral representation
(\ref{eq:w1}) of the Faddeeva function:
\begin{align}  \nonumber
i_n(\lambda,z_0) &= \frac{1}{\pi^{n/2}} \int\limits_0^{\infty} dz_1 \, e^{-z_1^2/4} \ldots \int\limits_0^{\infty} dz_n \, e^{-z_n^2/4} \\  \nonumber
& \times \delta(z_1+\ldots+z_n - \lambda) \, e^{-z_0(z_1+\ldots+z_n)}   \\
& = e^{-z_0\lambda}  \,i_n(\lambda,0).
\end{align}  
We are left therefore with the asymptotic analysis of
$i_n(\lambda,0)$.

One trivially gets $i_1(\lambda,0) = e^{-\lambda^2/4}/\sqrt{\pi}$.  In
general, one has to integrate over the cross-section of the hyperplane
$z_1 + \ldots + z_n = \lambda$ with the first (hyper-)octant $\R_+^n$.
In the limit $\lambda\to\infty$, the dominant contribution comes from
the vicinity of the point $(\lambda,\ldots,\lambda)/n$ of that
cross-section that is the closest to the origin.  One can therefore
introduce new coordinates centered at this point and oriented with
this cross-section.  For instance, for $n = 2$, one uses $z_1 =
\lambda/2 + r/\sqrt{2}$ and $z_2 = \lambda/2 - r/\sqrt{2}$ to
write
\begin{align*}
i_2(\lambda,0) & = \frac{1}{\pi} \int\limits_{-\lambda/\sqrt{2}}^{\lambda/\sqrt{2}} \frac{dr}{\sqrt{2}} e^{-\lambda^2/8 - r^2/4} \\
& = e^{-\lambda^2/8} \frac{\sqrt{2} \, \erf(\lambda/\sqrt{8})}{\sqrt{\pi}} \,.
\end{align*}
As $\lambda\to\infty$, the limits of the above integral can be
extended to infinity to get $i_2(\lambda,0) \simeq
\frac{\sqrt{2}}{\sqrt{\pi}} e^{-\lambda^2/8}$.  

Similarly, for $n = 3$, we use the polar coordinates $(r,\theta)$ in
the cross-section
\begin{align*}
z_1 & = \frac{\lambda}{3} + r\biggl(\frac{\cos\theta}{\sqrt{2}} + \frac{\sin\theta}{\sqrt{6}}\biggr), \\
z_2 & = \frac{\lambda}{3} + r\biggl(- \frac{2\sin\theta}{\sqrt{6}}\biggr), \\
z_3 & = \frac{\lambda}{3} + r\biggl(- \frac{\cos\theta}{\sqrt{2}} + \frac{\sin\theta}{\sqrt{6}}\biggr) ,
\end{align*}
such that $z_1+z_2+z_3= \lambda$.  As a consequence, we get
$z_1^2+z_2^2+z_3^2 = \lambda^2/3 + r^2$, from which
\begin{align*}
i_3(\lambda,0) & \approx \frac{2\pi}{\pi^{3/2}} \int\limits_{0}^{\infty} \frac{dr \, r}{\sqrt{3}} e^{-\lambda^2/12 - r^2/4} \\
& = e^{-\lambda^2/12} \frac{4}{\sqrt{3\pi}} \,.
\end{align*}
In general, we obtain
\begin{align}  \nonumber
i_n(\lambda,0) & \approx \frac{\omega_{n-1}}{\pi^{n/2}} \int\limits_{0}^{\infty} \frac{dr \, r^{n-2}}{\sqrt{n}} e^{-\lambda^2/(4n) - r^2/4} \\
\label{eq:IN_asympt}
& = e^{-\lambda^2/(4n)} \frac{2^{n-1}}{\sqrt{\pi n}} \,,
\end{align}
where $\omega_d = 2\pi^{d/2}/\Gamma(d/2)$ is the area of the unit
$d$-dimensional ball.  Substituting this asymptotic relation into
Eq. (\ref{eq:IN_z01}), we get the large-$\lambda$ behavior:
\begin{equation}  \label{eq:IN_asympt1}
I_N(\lambda,z_0) \approx \sum\limits_{n=1}^N \binom{N}{n} [\erf(z_0)]^{N-n} \, \frac{2^{n-1}}{\sqrt{\pi n}} \, e^{-(nz_0 + \lambda/2)^2/n}  \,.
\end{equation}
When $\lambda \gg Nz_0$, the dominant contribution comes from the term
with $n = N$ so that
\begin{equation}  \label{eq:IN_asympt2}
I_N(\lambda,z_0) \approx \frac{2^{N-1}}{\sqrt{\pi N}} \, e^{-(Nz_0 + \lambda/2)^2/N}  \,.
\end{equation}
In particular, one has
\begin{equation}  \label{eq:IN_asympt3}
I_N(\lambda,0) \approx \frac{2^{N-1}}{\sqrt{\pi N}} \, e^{-\lambda^2/(4N)}  \,.
\end{equation}

\section{Diffusion outside a ball}
\label{sec:ball}

In this Appendix, we consider another emblematic example of diffusion
in the exterior of a ball of radius $\rrho$: $\Omega = \{ \x\in\R^3
~:~ |\x|>\rrho\}$.

\subsection{Reminder for a single particle}

For the case of partially reactive boundary, the survival probability
reads \cite{Collins49}
\begin{eqnarray}
\nonumber
&& S_q(t|r_0) = 1 - \frac{\rrho \exp\bigl(-\frac{(r_0-\rrho)^2}{4Dt}\bigr)}{r_0(1 + 1/(q\rrho))} 
\biggl\{ \erfcx\biggl(\frac{r_0-\rrho}{\sqrt{4Dt}}\biggr) \\  
\label{eq:Sq_3d}
&& - \erfcx\biggl(\frac{r_0-\rrho}{\sqrt{4Dt}} + \left(1 + q\rrho\right) \frac{\sqrt{Dt}}{\rrho}\biggr)\biggr\} \,,
\end{eqnarray}
where $r_0 = |\x_0| \geq \rrho$ is the radial coordinate of the
starting point $\x_0$.  As diffusion is transient, the particle can
escape to infinity with a finite probability:
\begin{equation}  \label{eq:Sq_tinf_3d}
S_q(t|r_0) \xrightarrow[t\to\infty]{} S_q(\infty|r_0) = 1 - \frac{\rrho/r_0}{1 + 1/(q\rrho)} > 0.
\end{equation}

Expanding Eq. (\ref{eq:Sq_3d}) in a power series of $1/\sqrt{Dt}$ up
to the leading term, one gets the long-time behavior
\begin{equation}  
S_q(t|r_0) = S_q(\infty|r_0) + t^{-\alpha} \psi_q(r_0) + O(t^{-1}) ,
\end{equation}
with $\alpha = 1/2$ and
\begin{equation}  \label{eq:psiq_3d}
\psi_q(r_0) = \frac{q\rrho^2/r_0}{1+q\rrho}  \, \frac{r_0 - \rrho + \rrho/(1 + q\rrho)}{\sqrt{\pi D}} \,.
\end{equation}
This domain belongs therefore to class IV according to our
classification (\ref{eq:Sq_asympt}).

The probability density of the first-passage time, $H_q(t|r_0) =
-\partial_t S_q(t|r_0)$, follows immediately (see also
\cite{Grebenkov18}):
\begin{align}  
H_q(t|r_0) & = \frac{qD}{r_0} e^{-(r_0-\rrho)^2/(4Dt)} \biggl\{ \frac{\rrho}{\sqrt{\pi Dt}} \\   \nonumber
& - (1 + q\rrho) \erfcx\biggl(\frac{r_0-\rrho}{\sqrt{4Dt}} + (1+q\rrho) \frac{\sqrt{Dt}}{\rrho}\biggr) \biggr\}.
\end{align}
For a perfectly reactive target, one retrieves the Smoluchowski
result:
\begin{eqnarray}
S_\infty(t|r_0) &=& 1 - \frac{\rrho}{r_0} \erfc\biggl(\frac{r_0-\rrho}{\sqrt{4Dt}}\biggr), \\
H_\infty(t|r_0) &=& \frac{\rrho}{r_0} \, \frac{r_0-\rrho}{\sqrt{4\pi Dt^3}} \, e^{-(r_0-\rrho)^2/(4Dt)} .
\end{eqnarray}
In turn, the probability density $U_1(\ell,t|r_0)$ reads
\cite{Grebenkov20c}
\begin{equation} \label{eq:U1_3d_bis}
U_1(\ell,t|r_0) = \frac{\rrho \, e^{-\ell/\rrho}}{r_0} \, \frac{r_0-\rrho+\ell}{\sqrt{4\pi Dt^3}} e^{-(r_0 - \rrho +\ell)^2/(4Dt)} .
\end{equation}
This is a rare example when the probability density $U_1(\ell,t|\x_0)$
is found in a simple closed form.  Setting $\ell = 0$, one retrieves
the probability density of the first-passage time for a perfectly
absorbing sphere \cite{Smoluchowski17}.  Integrating the probability
density over $t$, one gets
\begin{equation}   \label{eq:Q1_3d}
Q_1(\ell,t|r_0) = \frac{\rrho \, e^{-\ell/\rrho}}{r_0} \erfc\biggl(\frac{r_0 - \rrho + \ell}{\sqrt{4Dt}}\biggr)  ,
\end{equation}
whereas the derivative with respect to $\ell$ yields the continuous
part of the probability density $\rho_1(\ell,t|\x_0)$:
\begin{align} 
& \rho_1(\ell,t|r_0)  = \biggl(1 - \frac{\rrho}{r_0} \erfc\biggl(\frac{r_0-\rrho}{\sqrt{4Dt}}\biggr) \biggr) \delta(\ell) \\  \nonumber
& + \frac{e^{-\ell/\rrho}}{r_0} \biggl( \erf\biggl(\frac{r_0 - \rrho + \ell}{\sqrt{4Dt}}\biggr) 
+ \frac{\rrho \, e^{-(r_0 - \rrho +\ell)^2/(4Dt)}}{\sqrt{\pi Dt}}   \biggr) 
\end{align}
(here we added explicitly the first term to account for the atom of
the probability measure at $\ell = 0$).  As diffusion is transient,
the crossing probability is below $1$:
\begin{equation}  \label{eq:Q1_ball}  
Q_1(\ell,\infty|r_0) = \int\limits_0^\infty dt \, U_1(\ell,t|r_0) = \frac{\rrho \, e^{-\ell/\rrho}}{r_0} < 1.
\end{equation}
In other words, the density $U_1(\ell,t|r_0)$ is not normalized to $1$
because the diffusing particle can escape to infinity before its
boundary local time has reached the threshold $\ell$.  Expectedly, the
mean first-crossing time is infinite, whereas the most probable
first-crossing time, corresponding to the maximum of
$U_1(\ell,t|r_0)$, is
\begin{equation}  \label{eq:Tmp_3d}
t_{\rm mp,1} = \frac{(r_0 - \rrho+\ell)^2}{6D} \,.
\end{equation}

\subsection{The crossing probability}

For the case of $N$ particles, we start by analyzing the crossing
probability $Q_N(\ell,\infty|r_0)$.  Rewriting
Eq. (\ref{eq:Sq_tinf_3d}) as
\begin{equation}  
S_q(\infty|r_0) = 1 - \rrho/r_0 + \frac{\rrho/r_0}{1 + q\rrho}  \,,
\end{equation}
and substituting it into Eq. (\ref{eq:UN_LaplaceP}), one gets
\begin{equation}   \label{eq:QN_3d_0}
Q_N(\ell,\infty|r_0)  = 1 - \L_{q,\ell}^{-1} \left\{ \frac{ \bigl[1 - \rrho/r_0 + \frac{\rrho/r_0}{1 + q\rrho} \bigr]^N}{q} \right\} .
\end{equation}
Using the binomial expansion and the identity
\begin{equation}  \label{eq:IL_identity}
\L_{q,\ell}^{-1} \biggl\{ \frac{1}{q(1+qR)^n}\biggr\} = 1 - e^{-\ell/R}\sum\limits_{k=0}^{n-1} \frac{(\ell/R)^k}{k!} \,,
\end{equation}
we evaluate the inverse Laplace transform of each term that yields
after re-arrangment of terms:
\begin{align} \nonumber
Q_N(\ell,\infty|r_0) &= e^{-\ell/\rrho} \sum\limits_{k=0}^{N-1} \frac{(\ell/\rrho)^k}{k!}  \\
& \times \biggl(1 - \sum\limits_{n=0}^k \binom{N}{n} \alpha^n (1-\alpha)^{N-n} \biggr),
\end{align}
with $\alpha = \rrho/r_0$.  For $N = 1$, we retrieve
Eq. (\ref{eq:Q1_ball}).  At $r_0 = \rrho$, one gets a simpler relation
\begin{equation} \label{eq:QNinf_3d_r1}
Q_N(\ell,\infty|\rrho) = e^{-\ell/\rrho} \sum\limits_{k=0}^{N-1} \frac{(\ell/\rrho)^k}{k!} \,.
\end{equation}
For a fixed $\ell/\rrho$ and large $N$, one has
\begin{equation}  \label{eq:QNinf_3d_r1_N}
Q_N(\ell,\infty|\rrho) \simeq 1 - \frac{(\ell/\rrho)^N e^{-\ell/\rrho}}{N!}  \qquad (N\to \infty),
\end{equation}
i.e., the crossing probability rapidly approaches $1$.

\begin{figure}
\begin{center}
\includegraphics[width=88mm]{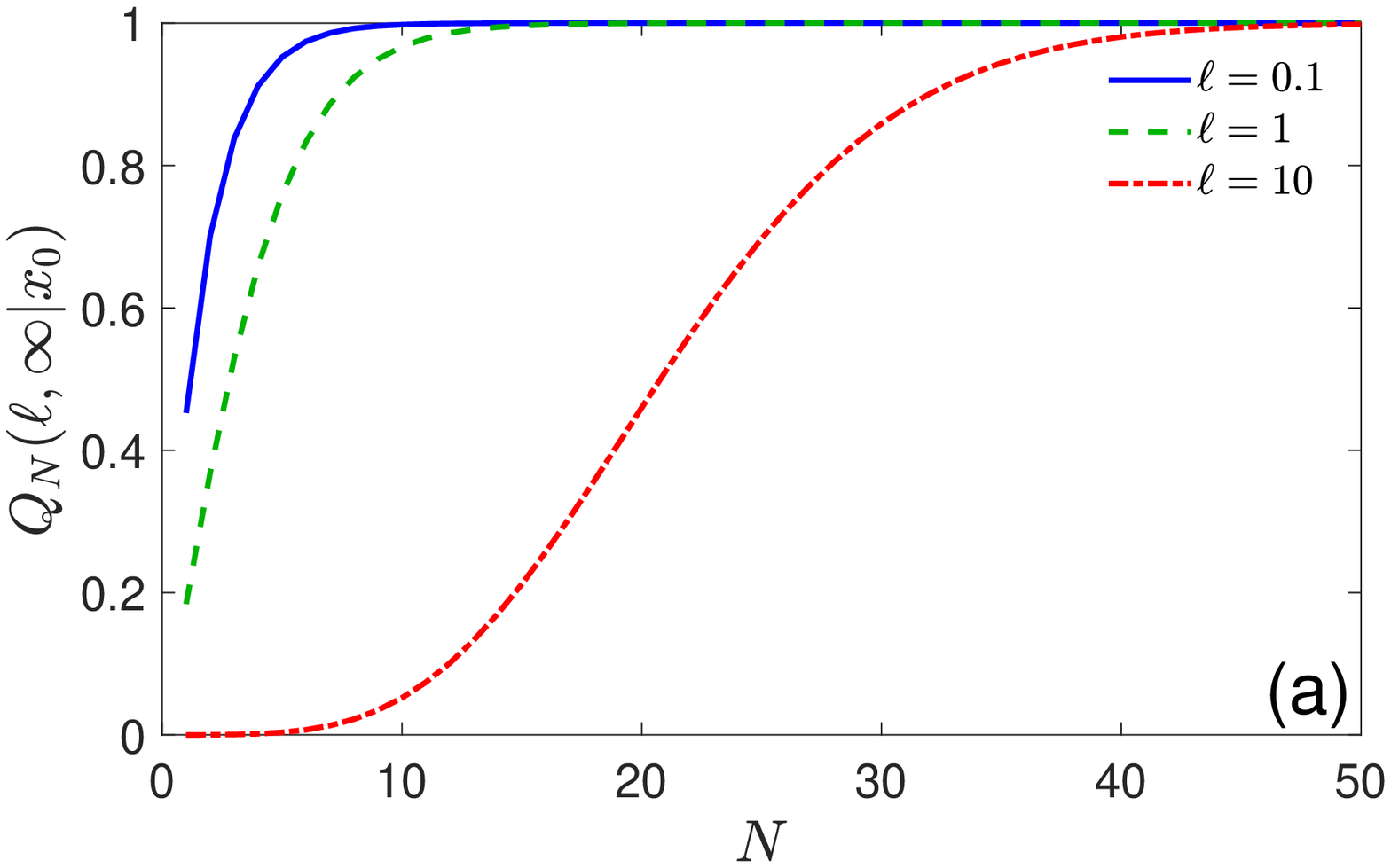} % QNinf_3d_r2.eps}
\includegraphics[width=88mm]{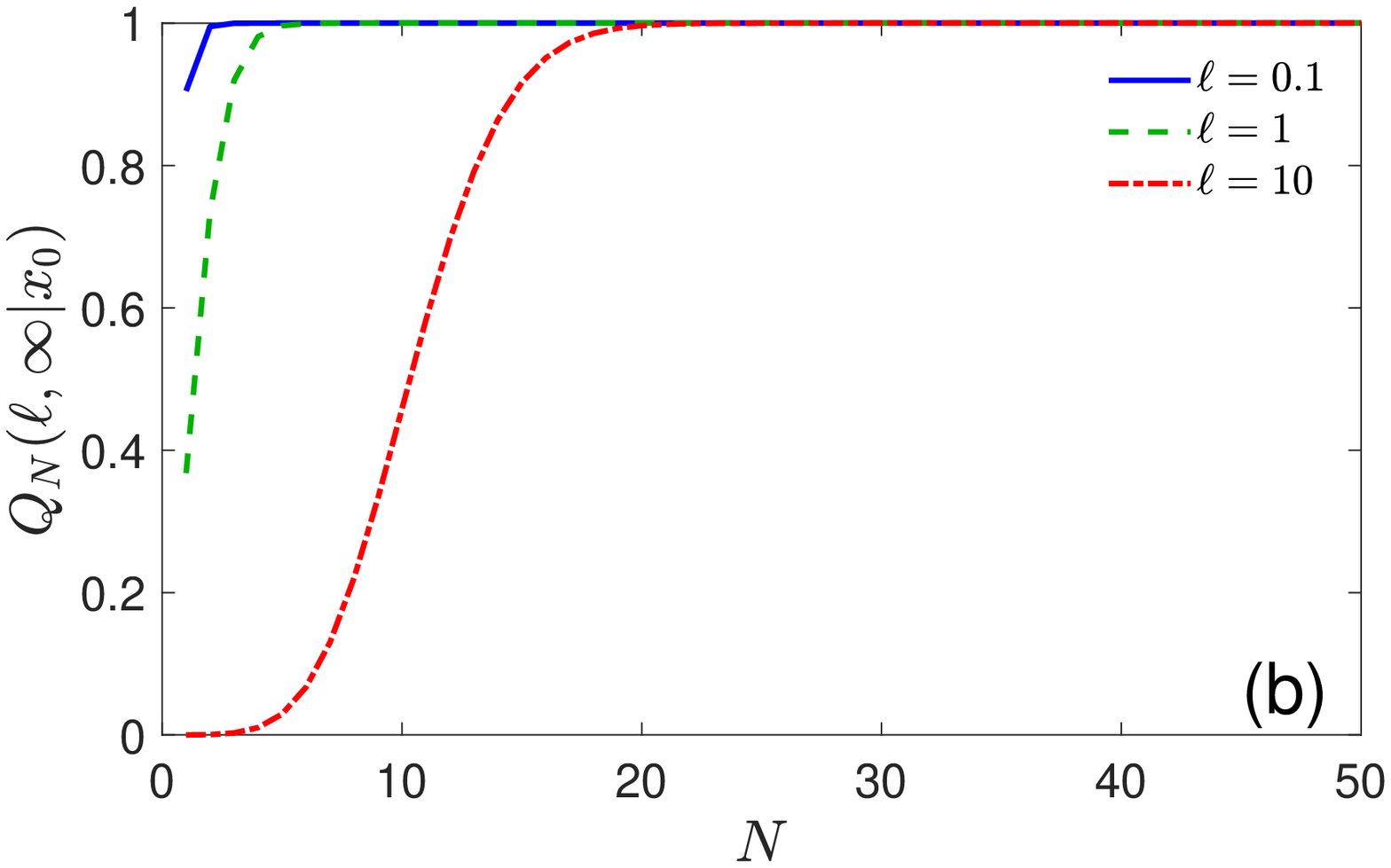} % QNinf_3d_r1.eps}
\end{center}
\caption{
Crossing probability $Q_N(\ell,\infty|\x_0)$ for $N$ particles
diffusing in the exterior of a ball of radius $R = 1$, with three
values of $\ell$ indicated in the legend, and $|\x_0| = 2R$ {\bf (a)}
and $|\x_0| = R$ {\bf (b)}. }
\label{fig:QNinf_3d}
% [QN] = A_localtime6_QNinf_3d_fig(2);
% [QN] = A_localtime6_QNinf_3d_fig(1);
\end{figure}

Figure \ref{fig:QNinf_3d} illustrates the behavior of the crossing
probability $Q_N(\ell,\infty|r_0)$ as a function of $N$.  One sees
that $Q_N(\ell,\infty|r_0)$ monotonously grows with $N$ and rapidly
approaches $1$, whereas the threshold $\ell$ and the starting point
$\x_0$ determine how fast this limit is reached.

\subsection{PDF of the total boundary local time}

Setting $z_0 = (r_0 - \rrho)/\sqrt{4Dt}$ and $\alpha = \rrho/r_0$, one
can rewrite the survival probability from Eq. (\ref{eq:Sq_3d}) as
\begin{align} \nonumber
S_q(t|r_0) &= 1 - \frac{\alpha}{1 + 1/(q\rrho)} + \frac{\alpha}{1 + 1/(q\rrho)}  \\
& \times \biggl(\erf(z_0) + e^{-z_0^2} \erfcx\bigl(z'_0 + q \sqrt{Dt}\bigr) \biggr),
\end{align}
where $z'_0 = z_0 + \sqrt{Dt}/\rrho$, and the expression in
parentheses resembles the survival probability from
Eq. (\ref{eq:Sq_1d}) for diffusion on the half-line.
The probability density of the total boundary local time $\ell_t$
reads then
\begin{equation}
\rho_N(\ell,t|r_0) = \bigl(S_\infty(t|r_0)\bigr)^N \delta(\ell) + \frac{I_N^{3d}\bigl(\ell/\sqrt{Dt}, z_0)}{\sqrt{Dt}}  \,,
\end{equation}
where $S_\infty(t|r_0) = 1 - \alpha \,\erfc(z_0)$ and
\begin{align}  \nonumber
& I_N^{3d}(\lambda,z_0) = \int\limits_{-\infty}^\infty \frac{dq}{2\pi} e^{-iq\lambda} \biggl\{
\biggl[1 - \frac{\alpha}{1 + i/(qR')} \\  \nonumber
& + \frac{\alpha}{1 + i/(qR')} \biggl(\erf(z_0) + e^{-z_0^2} \erfcx\bigl(z_0 + 1/R' - iq \bigr) \biggr) \biggr]^N \\
& - \bigl[1 - \alpha \,\erfc(z_0)\bigr]^N \biggr\} ,
\end{align}
with $R' = R/\sqrt{Dt}$.  We skip the analysis of this function and
the consequent asymptotic behavior for $\rho_N(\ell,t|r_0)$, see
Appendix \ref{sec:1D_rhoN} for a similar treatment for diffusion on
the half-line.

\subsection{PDF of the first-crossing time}

Substituting Eq. (\ref{eq:Sq_3d}) into Eq. (\ref{eq:UN_Fourier}), one
gets
\begin{align} \nonumber
U_N(\ell,t|r_0) & = \frac{ND e^{-z_0^2}}{r_0 \sqrt{Dt}} \int\limits_{-\infty}^\infty \frac{dq}{2\pi} e^{iq\ell/\sqrt{Dt}} 
\biggl[1 - \frac{\alpha \, \erfc(z_0)}{1 - i/(q\rrho')} \\  \nonumber
& + \frac{\alpha \, e^{-z_0^2}\erfcx(z_0 + R' + iq)}{1 - i/(q\rrho')}\biggr]^{N-1} \\   \label{eq:UN_3d}
& \times \biggl(\frac{\rrho'}{\sqrt{\pi}} - (1 + iq\rrho') \erfcx(z_0 + R' + iq)\biggr),
\end{align}
where $\rrho' = \rrho/\sqrt{Dt}$.  The short-time behavior of this
function is given by Eq. (\ref{eq:UN_t0_x0}) for $|\x_0| > R$ and
Eq. (\ref{eq:UN_short}) for $|\x_0| = R$, respectively.

To get the long-time behavior from Eq. (\ref{eq:UN_asympt}), we need to
evaluate the following inverse Laplace transform
\begin{align}
\Psi_N(\x_0,\ell) & = \frac{R \alpha^N}{\sqrt{\pi D}} \\  \nonumber
& \times \L_{q,\ell}^{-1} \biggl\{ \frac{1}{q} \biggl(1 - \frac{1}{1+qR}\biggr) 
\biggl(\beta + \frac{1}{1+qR}\biggr)^N \biggr\} ,
\end{align}
where we used Eqs. (\ref{eq:Sq_tinf_3d}, \ref{eq:psiq_3d}), and set
$\beta = (1-\alpha)/\alpha$.  Using the binomial expansion and the
identity (\ref{eq:IL_identity}), we get after simplifications:
\begin{equation}  \label{eq:PsiN_3d}
\Psi_N(\x_0,\ell) = \frac{R e^{-\ell/R} }{\sqrt{\pi D}} \sum\limits_{n=0}^N \binom{N}{n} (1-R/r_0)^{N-n} \frac{(\ell/r_0)^n}{n!} \,.
\end{equation}
Substituting this expression into Eq. (\ref{eq:UN_asympt}), we obtain
\begin{equation}  \label{eq:UN_3d_tinf}
U_N(\ell,t|r_0) \simeq \frac{NR \,e^{-\ell/R} }{\sqrt{4\pi Dt^3}} \sum\limits_{n=0}^N \binom{N}{n} (1-R/r_0)^{N-n} \frac{(\ell/r_0)^n}{n!} \,.
\end{equation}
In the particular case $r_0 = R$, the above sum is reduced to a single
term with $n = N$ so that
\begin{equation}
U_N(\ell,t|R) \simeq \frac{R\, e^{-\ell/R} }{\sqrt{4\pi Dt^3}} \, \frac{(\ell/r_0)^N}{(N-1)!} \,.
\end{equation}
We conclude that, contrarily to the one-dimensional case, the
probability density $U_N(\ell,t|r_0)$ exhibits the same $t^{-3/2}$
asymptotic decay for any $N$, while the population size affects only
the prefactor.  In particular, the mean first-crossing time is always
infinite.

\begin{figure}
\begin{center}
\includegraphics[width=88mm]{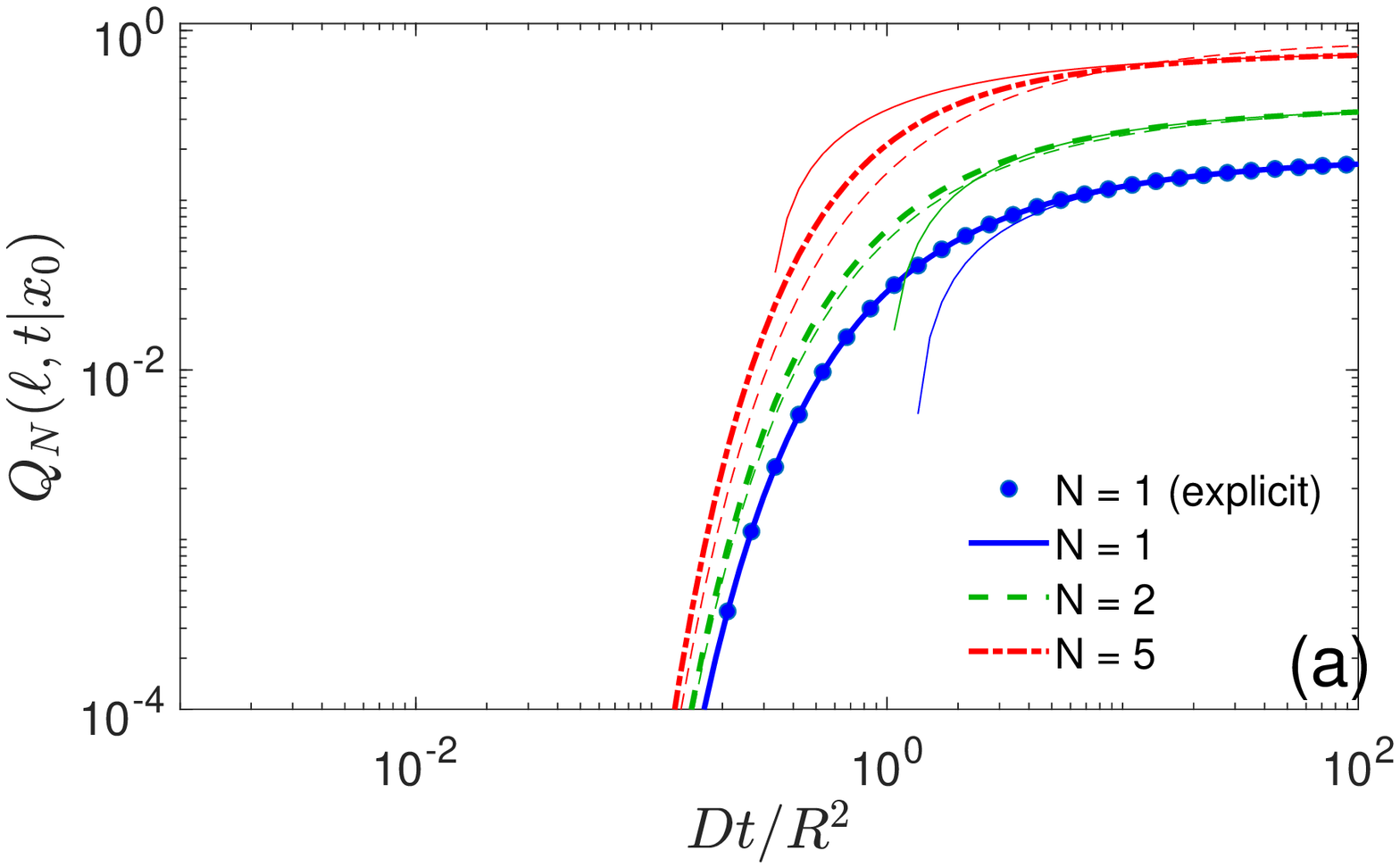} % QN_3d_r2new.eps}
\includegraphics[width=88mm]{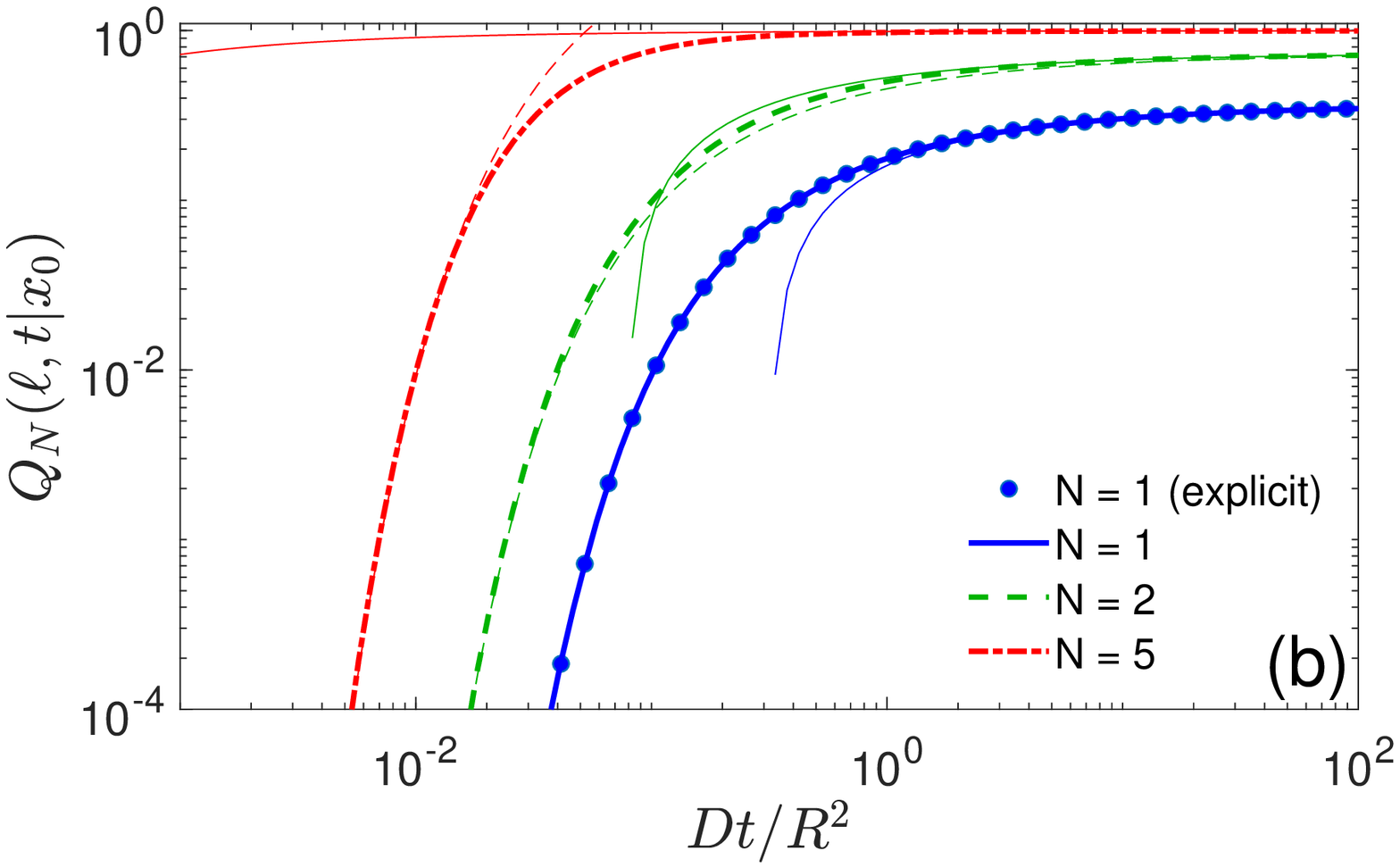} % QN_3d_r1.eps}
\end{center}
\caption{
Cumulative distribution function $Q_N(\ell,t|\x_0)$ of the
first-crossing time $\T_{\ell,N}$ for $N$ particles diffusing in the
exterior of a ball of radius $\rrho = 1$, with $\ell = 1$, $D = 1$,
$|\x_0| = 2$ {\bf (a)} and $|\x_0| = 1$ {\bf (b)}.  Symbols present
the explicit form (\ref{eq:Q1_3d}) for a single particle, whereas
thick lines show the numerical integration in
Eq. (\ref{eq:QN_Fourier}).  Thin lines indicate the long-time
asymptotic relation (\ref{eq:QN_asympt}), while thin dashed lines
present the short-time behavior in Eq. (\ref{eq:QN_short1}) for
$|\x_0| = 2$ and Eq. (\ref{eq:QN_short2}) for $|\x_0| = 1$,
respectively. }
\label{fig:QN_3d}
% [Q1,Q2,Q5, t] = A_localtime6_QN_3d_fig2(U1,U2,U5); % r0=2
% [Q1,Q2,Q5, t] = A_localtime6_QN_3d_fig(U1,U2,U5);  % r0=1
\end{figure}

Figure \ref{fig:UN_3d} illustrated the probability density
$U_N(\ell,t|r_0)$ and its asymptotic behavior for $\ell/\rrho = 1$.
%When the species start a distance away from the stock region (panel
%(a)), $U_N(\ell,t|r_0)$ looks as being just ``shifted'' upwards by
%increasing $N$, in agreement with the short-time behavior in
%Eq. (\ref{eq:UN_t0_x0}).  In particular, the most probable
%first-crossing time remains close to that of a single species.  As the
%species need first to reach the stock region, speed up of the
%depletion by having many species is modest.  The situation is
%drastically different when the species start on the stock region
%(panel (b)).  In this case, some species may stay close to the stock
%region, repeatedly returning to it and rapidly consuming its
%resources.  One sees that the total boundary local time reaches a
%prescribed threshold $\ell$ much faster, and the probability density
%$U_N(\ell,t|r_0)$ is shifted towards shorter times as $N$ increases.
%In both panels, the short-time and long-time asymptotic relations
%derived above are accurate.  We stress that the mean first-crossing
%time and higher-order moments are infinite and thus not informative
%here.
%Some other aspects of this depletion problem, such
%as the cumulative distribution function $Q_N(\ell,t|\x_0)$, the
%probability of depletion, and their dependence on $N$, are discussed
%in \secball.
To provide a complementary view onto the properties of the
first-crossing time, we also present the cumulative distribution
function $Q_N(\ell,t|r_0)$ on Fig. \ref{fig:QN_3d}.  As discussed
previously, when the particles are released on the stock region, the
stock depletion occurs much faster when $N$ increases.

\begin{figure}
\begin{center}
\includegraphics[width=88mm]{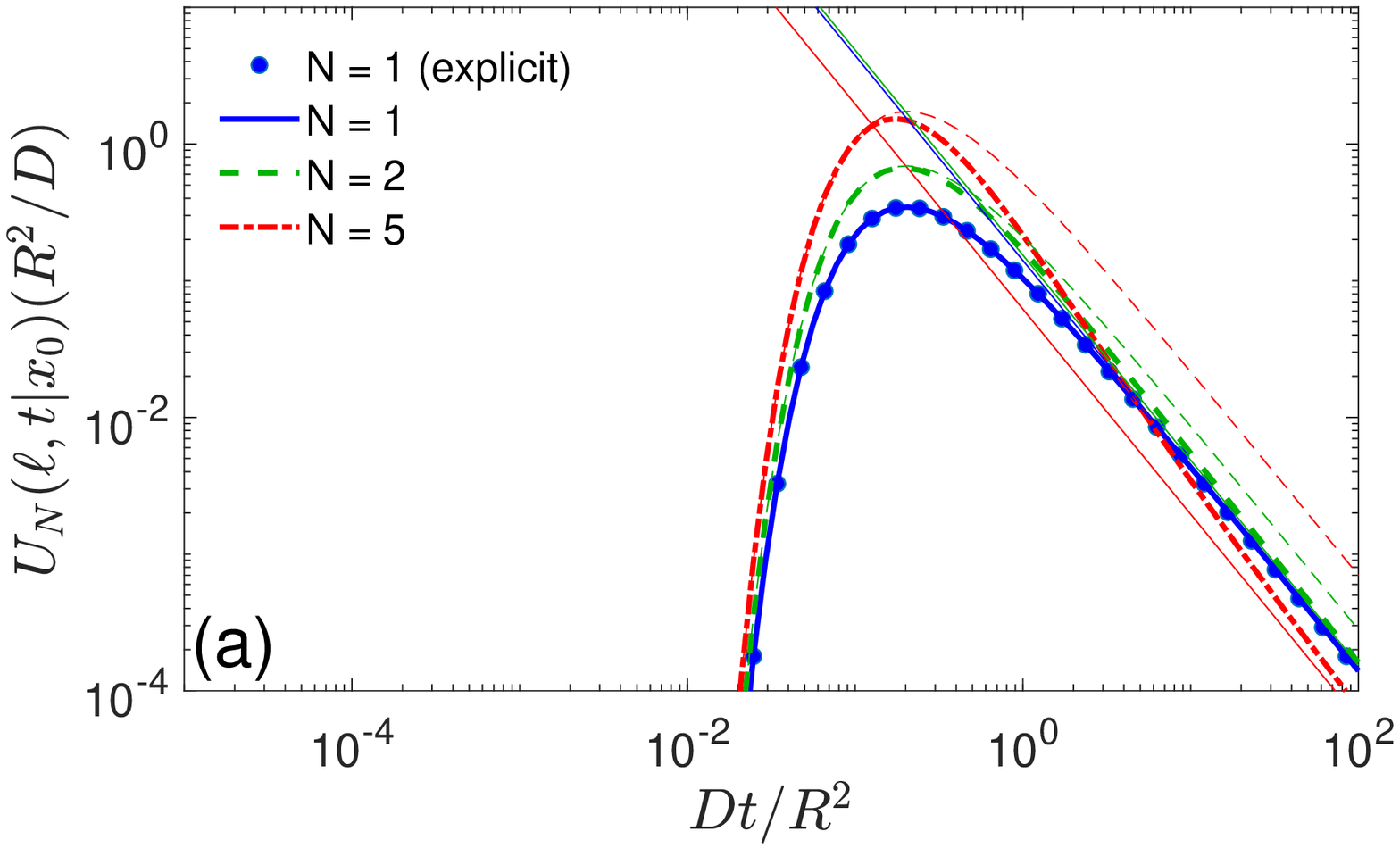} % UN_3d_r2_ell01.eps}
\includegraphics[width=88mm]{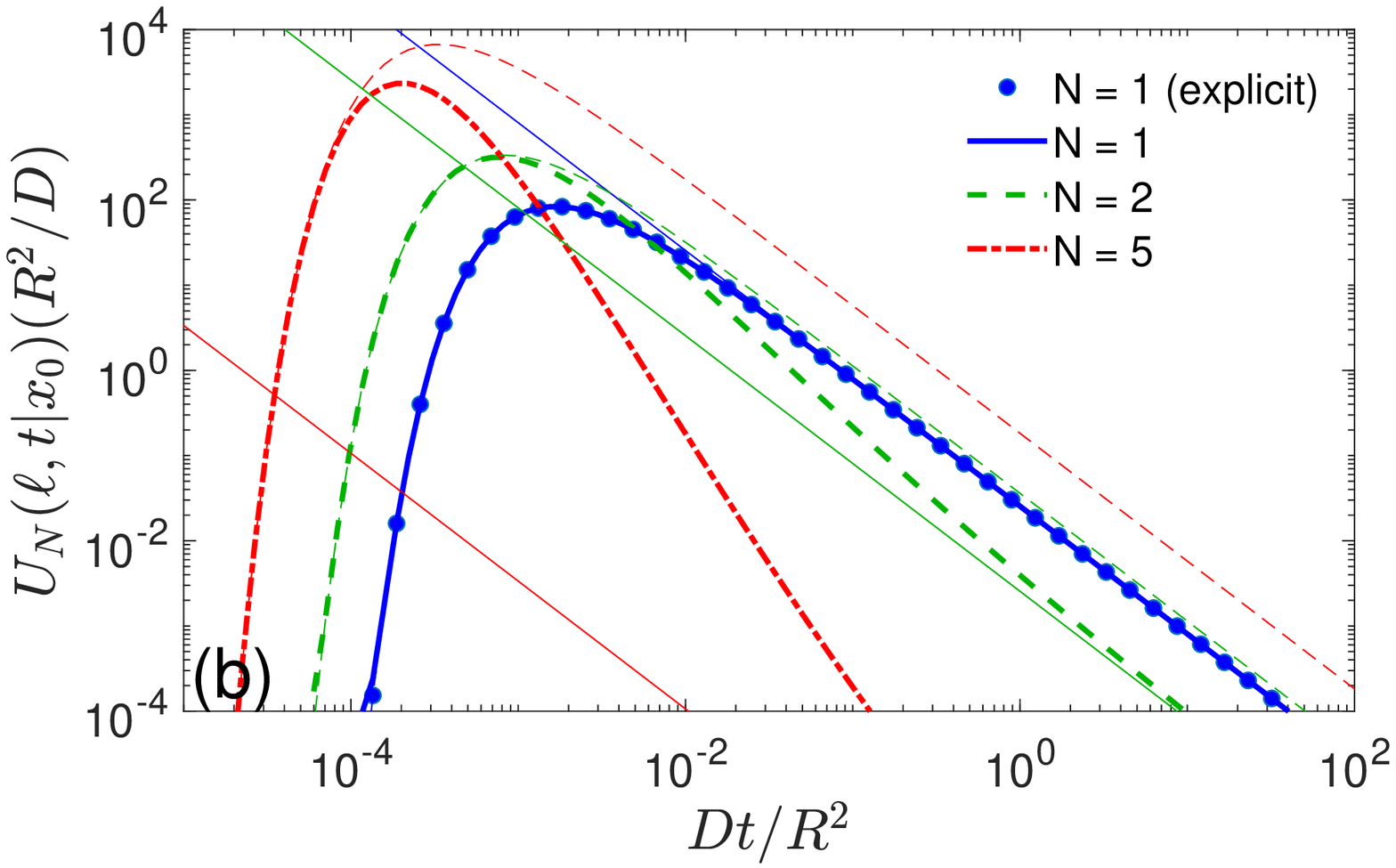} % UN_3d_r1_ell01.eps}
\end{center}
\caption{
Probability density function $U_N(\ell,t|\x_0)$ of the first-crossing
time for $N$ particles diffusing in the exterior of a ball of radius
$\rrho = 1$, with $\ell = 0.1$, $D = 1$, $|\x_0| = 2$ {\bf (a)} and
$|\x_0| = 1$ {\bf (b)}.  Symbols present the explicit form
(\ref{eq:U1_3d}) for a single particle, whereas thick lines show the
numerical integration in Eq. (\ref{eq:UN_3d}).  Thin lines indicate
the long-time asymptotic relation (\ref{eq:UN_3d_tinf}), while thin
dashed lines present the short-time behavior in
Eq. (\ref{eq:UN_t0_x0}) for $|\x_0| = 2$ and Eq. (\ref{eq:UN_short})
for $|\x_0| = 1$. }
\label{fig:UN_3d_ell01}
% [U1,U2,U5, t] = A_localtime6_UN_3d_fig2(U1,U2,U5); % r0=2
% [U1,U2,U5, t] = A_localtime6_UN_3d_fig(U1,U2,U5);  % r0=1
\end{figure}

For comparison, we also consider a smaller threshold $\ell/\rrho =
0.1$, for which the probability density $U_N(t,\ell|r_0)$ is shown in
Fig. \ref{fig:UN_3d_ell01}.  As previously, the behavior strongly
depends on whether the particles start on the stock region (or close
to it) or not.  In the former case ($r_0 = \rrho$), the maximum of the
probability density for $\ell/\rrho = 0.1$ is further shifted to
smaller times, as expected.  Note also that $U_N(\ell,t|r_0)$ for $N =
5$ exhibits a transitory regime at intermediate times with a rapid
decay, so that the long-time behavior in Eq. (\ref{eq:UN_3d_tinf}),
which remains correct, is not much useful here, as it describes the
probability density of very small amplitude.
% (due to the factor $1/N!$).
%
In turn, for $r_0 = 2\rrho$, the three curves on
Fig. \ref{fig:UN_3d_ell01}(a) resemble those on
Fig. \ref{fig:UN_3d}(a), because the limiting factor here is finding
the stock region.  In particular, setting $\ell = 0$, one would get
the probability density of the fastest first-passage time to the
perfectly absorbing target
\cite{Weiss83,Basnayake19,Lawley20,Lawley20b,Lawley20c,Grebenkov20d}.

\section{Numerical computation}
\label{sec:numerical}

As a numerical computation of the inverse Laplace transform may be
unstable, it is convenient to replace the Laplace transform by the
Fourier transform.  This is equivalent to replacing the generating
function $\E_{\x_0}\{ e^{-q\ell_t}\}$ of $\ell_t$ by its
characteristic function $\E_{\x_0}\{ e^{iq\ell_t}\}$.  In this way, we
get
\begin{align*}  \nonumber
\rho_N(\ell,t|\x_0) & = \int\limits_{-\infty}^\infty \frac{dq}{2\pi} e^{-iq\ell} \, \E_{\x_0}\{ e^{iq\ell_t}\}  \\  \nonumber
& = \int\limits_{-\infty}^\infty \frac{dq}{2\pi} e^{-iq\ell}\, \bigl(\E_{\x_0}\{ e^{iq\ell_t^1}\}\bigr)^N \\ 
& = \int\limits_{-\infty}^\infty \frac{dq}{2\pi} e^{-iq\ell}\, \bigl( S_{-iq}(t|\x_0) \bigr)^N  .
\end{align*}
Since the survival probability $S_\infty(t|\x_0)$ is strictly positive
for any $\x_0 \notin \Gamma$, the total boundary local time $\ell_t$
can be zero with a finite probability $[S_\infty(t|\x_0)]^N$, and it
is convenient to subtract the contribution of this atom in the
probability measure explicitly, so that
\begin{align} \label{eq:rhoN_ell}
& \rho_N(\ell,t|\x_0) = \bigl( S_{\infty}(t|\x_0) \bigr)^N \delta(\ell) \\   \nonumber
& \quad + \int\limits_{-\infty}^\infty \frac{dq}{2\pi} e^{-iq\ell}\, \biggl[\bigl( S_{-iq}(t|\x_0) \bigr)^N - \bigl( S_{\infty}(t|\x_0) \bigr)^N \biggr] ,
\end{align}
where $\delta(\ell)$ is the Dirac distribution.  The probabilistic
interpretation of this relation is straightforward: as the total
boundary local time remains $0$ until the first arrival of any of the
particles onto the stock region, the random event $\ell_t = 0$
(expressed by $\delta(\ell)$) has a strictly positive probability
$\bigl( S_{\infty}(t|\x_0) \bigr)^N$, i.e., the probability that none
of $N$ particles has arrived onto the stock region up to time $t$.
Since the diffusion equation (\ref{eq:Sq_diff}) and the Robin boundary
condition (\ref{eq:Sq_BC_Robin}) are linear, one has
\begin{equation}
S_{iq}(t|\x_0) = S_{-iq}^*(t|\x_0),
\end{equation}
where asterisk denotes the complex conjugate.  As a consequence, one
can rewrite Eq. (\ref{eq:rhoN_ell}) as
\begin{align} \label{eq:rhoN_ell2}
& \rho_N(\ell,t|\x_0) = \bigl( S_{\infty}(t|\x_0) \bigr)^N \delta(\ell) \\   \nonumber
& \quad + \Re \biggl\{ \int\limits_0^\infty \frac{dq}{\pi} e^{iq\ell}
\biggl[\bigl( S_{iq}(t|\x_0) \bigr)^N - \bigl( S_{\infty}(t|\x_0) \bigr)^N \biggr] \biggr\} .
\end{align}

Similarly, the probability density $U_N(\ell,t|\x_0)$ and the
cumulative distribution function $Q_N(\ell,t|\x_0)$ can be written in
the Fourier form as
\begin{equation}   \label{eq:UN_Fourier}
U_{N}(\ell,t|\x_0) = \Re\biggl\{ \int\limits_0^\infty \frac{dq}{\pi} \, \frac{e^{iq\ell}}{iq} \, 
\biggl(- \partial_t [S_{iq}(t|\x_0)]^N \biggr) \biggr\}
\end{equation}
and
\begin{equation}   \label{eq:QN_Fourier}
Q_{N}(\ell,t|\x_0) = \Re\biggl\{ \int\limits_0^\infty \frac{dq}{\pi} \, \frac{e^{iq\ell}}{iq} \, \biggl([S_{iq}(t|\x_0)]^N - 1 \biggr)\biggr\} .
\end{equation}

\end{document}